\documentclass[prb,rsi,amsmath,amssymb,reprint]{revtex4-1}

\usepackage{graphicx}
\usepackage{dcolumn}
\usepackage{bm}
\usepackage{natbib}
\setcitestyle{square}
\usepackage{soul}

\usepackage{amsmath}
\usepackage{braket}
\usepackage{mathrsfs}
\usepackage{amsfonts}
\usepackage{color}
\usepackage[normalem]{ulem} 

\usepackage{hyperref}

\hypersetup{
    colorlinks=true,
    linkcolor=blue,
    filecolor=magenta,      
    urlcolor=black,
    citecolor=red,
}

\begin{document}
\title{Machine Learning S-wave Scattering Phase Shifts Bypassing the Radial Schr\"{o}dinger Equation}

\author{Alessandro Romualdi}
\email{alessandro.romualdi@yahoo.com}
\affiliation{%
}%
\author{Gionni Marchetti}
\email{gionnimarchetti@ub.edu}
\affiliation{%
Departament de F\'{i}sica de la Mat\`{e}ria Condensada, Universitat de Barcelona, Carrer Mart\'{i} i Franqu\`{e}s 1, 08028, Barcelona, Spain\\
}%

\affiliation{%
National Institute of Chemical Physics and Biophysics, R{\"a}vala 10,  10143, Tallinn, Estonia\\
}%

\date{\today}

\begin{abstract}
We present a \emph{proof of concept} machine learning model resting on a convolutional neural network capable to yield accurate scattering s-wave phase shifts caused by different three-dimensional spherically symmetric potentials at fixed collision energy thereby bypassing the radial Schr\"{o}dinger equation. 
In out work, we discuss how the Hamiltonian can serve as a guiding principle in the construction of a physically-motivated descriptor.
The good performance, even in presence of bound states in the data sets, exhibited by our model that accordingly is trained on the Hamiltonian through each scattering potential, demonstrates the feasibility of this proof of principle. 
 \end{abstract}

\maketitle


Phase shifts play a key role in the partial wave analysis, and are the crucial quantities for obtaining the scattering amplitude by which the elastic scattering properties of the physical or chemical system of interest can be calculated~\cite{joachain1975, piel2020}. In particular, phase shifts $\delta_0$ in s-wave, i.e., when the angular momentum number $l=0$, are significant for atomic collisions in ultracold regime where the s-wave scattering channel dominates~\cite{fermi1936, huang1957, idziaszek2006}. Moreover, in Coulomb systems the  s-wave phase shift  allows the $\delta_l$ with $l=1,2, \cdots$ to be recursively determined starting from the knowledge of $\delta_0$~\cite{pain2018}. 

Therefore their accurate computation is in great demand in the various fields of science. In this regard it is worth noting that Kohn developed one of the most used method for accurately computing the phase shifts based on the variational principle and the Schr\"{o}dinger equation~\cite{kohn1948, nesbet1968, joachain1975}, and very recently a new method and new computer code, based on a recurrence relation and the variable phase method (VPM)~\cite{calogero1963, calogero1967, babikov1967, fano1986}, respectively, aiming at the same task have been put forward~\cite{ pain2018, palov2021}.

\begin{figure}[htp]
\resizebox{0.40\textwidth}{!}{%
  \includegraphics{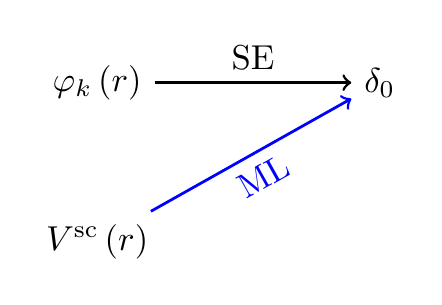}
}
\caption{A Schematic view of our \emph{proof of concept}. Our ML model is trained on the Hamiltonian, through the scattering potential  $V^{\rm sc}\left(r\right) $ and not on the phase function $\varphi_k\left(r\right)$ in the real space, for learning the scattering phase shifts $\delta_0$ at a given wave number $k$. Instead, the standard approach is to numerically integrate the  radial Schr\"{o}dinger equation  for yielding  $\varphi_k\left(r\right)$ and the corresponding asymptotic  $\delta_0$ with initial condition $\delta_0\left(0\right)=0$ at the origin, i.e. at $r=0$.}
\label{fig:cartoonProof}
\end{figure}

While the traditional numerical methods such as the one presented in Ref.~\cite{palov2021} are based on the  algorithms which numerically integrate the  Schr\"{o}dinger equation for a given scattering potential  $V^{\rm sc} $, we herein propose a supervised machine learning (ML) approach for yielding accurate phase shifts through the empirical fitting of data, therefore bypassing the radial Schr\"{o}dinger equation (SE). Although our work is, to some extent, motivated by the recent advances of ML and its applications to physics, chemistry and materials science~\cite{ zdeborova2017, carleo2019, schmidt2019} which prove successful in yielding accurate predictions of physical quantities, such as the molecular ground state energy \cite{faber2017, bartok2017}, polarizability~\cite{montavon2013}, atomization energy~\cite{rupp2012}, density of states at Fermi energy of crystalline solids~\cite{schuett2014}, identification of the phases of matter~\cite{wang2016, carrasquilla2017, vanNieuwenburg2018}, development of ML potentials~\cite{behler2007, bartok2010, thompson2015, shapeev2016, behler2016} and construction of deep-learning wavefunctions of the electronic Schr\"{o}dinger equation~\cite{hermann2020,manzhos2020}, its purpose is essentially different being a \emph{proof of concept} (see Fig.~\ref{fig:cartoonProof}). Indeed, in this paper we wish to address the crucial role played by the Hamiltonian $H$ for constructing  a physically motivated descriptor (or fingerprint)~\cite{ghiringhelli2015, bartok2013, rossi2020, musil2021}. First, the Hamiltonian contains all the possible  physical information about the system under scrutiny~\cite{gross1996} e.g., the symmetries such as the translations and rotations under which it is invariant. Second, the physical scale over which the  phenomena occur  is naturally provided to the descriptor without the need to  heuristically  determine  \emph{ad hoc} cut-offs as routinely done for the atomistic systems
~\cite{behler2016, musil2021}.

To support the above observations, first  we shall  limit ourselves to the regression task that involve the prediction of the quantum phase shifts $\delta_0$ at a fixed collision energy in  the presence of a three-dimensional spherically symmetric regular potential. In such a case the Hamiltonian reads  $H = T +  V^{\rm sc} $, where $T$ and $V^{\rm sc}$ are the kinetic energy and the scattering potential, respectively. Therefore the scattering potential $V^{\rm sc}$  uniquely determines the Hamiltonian $H$~\cite{hohenberg1964}. This situation is analogous to that thereby the Coulomb Matrix (CM) descriptor~\cite{rupp2012, lilienfeld2013} is built on. Indeed, the CM's entries correspond to the Coulomb terms of the external potential in  the molecular Hamiltonian. 

Second, our ML model consists in a convolutional neural network (CNN) trained on the Hamiltonian, through the scattering potential $V^{\rm sc}$ and not on the phase function $\varphi_k$ that can be obtained only by solving the radial  Schr\"{o}dinger equation (see Fig.~\ref{fig:cartoonProof}). Accordingly, the regression task for the phase shifts will be performed by feeding images $X$ of $V^{\rm sc}$ instances together with their labels $\delta_0$ to the CNN. It is worth noting  that our  machine model  is somehow equivalent to that employed by Mills et al.~\cite{mills2018}  to predict ground-state energy of an electron in confining two-dimensional electrostatic potentials.
 
Overall, our machine learning model is capable of predicting the target phase shifts, caused by different three-dimensional spherically symmetric regular potentials at fixed collision energy with relative error much less than $1\%$, and corresponding ML accuracy in terms of the average of the mean absolute percentage error (MAPE) $\lesssim 0.1\%$. This is remarkable, considering that the approximate phase shifts computed in the first Born approximation and the exact ones are considered to agree very well when the relative error is $\lesssim 5\%$~\cite{joachain1975}.

Although  our findings are a consequence of the universal approximation theorem for the neural networks \cite{yarotsky2018} which  suggests that the continuous mapping between the scattering potential $V^{\rm sc}\left(r\right)$ in the real space and the corresponding phase shift $\delta_0\left(r\right)$ can be learned by means of a feed-forward CNN \cite{hubel1962, fukushima1980}, similar to that successfully used in image recognition tasks \cite{lecun1998, krizhevsky2012, szegedy2015, silver2016, mallat2016, lin2017}, they do support the idea that the Hamiltonian of any given system should serve as a guiding principle in the search for physically-motivated descriptors for ML applications to physical sciences.

Note that throughout this paper the computations were performed assuming the typical scattering units, that is, $\hbar=2 m=1$, $m$ being the reduced mass of the scattering particle. Therefore  in such units $k$, $k^{2}$ are the wave number and energy of the scattering particle in the center of mass reference frame, respectively.

\section{Methods} \label{methods}

\subsection{Three-Dimensional Spherically Symmetric  Potentials} \label{potentials}

In order to address the problem of learning quantum scattering phase shifts from data, we consider three-dimensional spherically symmetric potentials that satisfy some nice mathematical properties in accordance with the scattering theory. In particular, the latter requires the scattering potentials to be reasonable smooth functions of the interparticle distance and to fall off sufficiently fast in the asymptotic region \cite{calogero1967}.

In the present work our choice of the potentials was dictated by their theoretical and practical importance in various branches of physics and chemical physics. In the following we briefly list the potentials under scrutiny.

 The Thomas-Fermi (TF) potential (or Yukawa potential)  $V^{\rm TF}$ accounts for the screened Coulomb interaction and the strong interaction in condensed matter systems~\cite{ashcroft1976,  meyer1981, giuliani2005} and nuclear and particle physics~\cite{povh2002}, respectively.
In its attractive form this potential has the following expression $V^{\rm TF}\left(r\right) = - V_{0}^{\rm TF} \exp\left(-qr\right)/r$ where the positive quantities $r, q, V_{0}^{\rm TF} $  denote the interparticle distance, the screening parameter and the potential's strength, respectively.
 
A linear combination of the TF potentials gives rise to a scattering potential of significant importance in nuclear and condensed matter physics~\cite{ bali1967, nagy1998}. Here we shall consider a double Yukawa (DY) potential $V^{\rm DY}$ which is just a linear combination of two TF potentials, i.e. $V^{\rm DY}\left(r\right)  = - V_{\rm 01} \exp\left(-q_1r\right)/r + V_{\rm 02} \exp\left(-q_2r\right)/r $. In this paper we shall tune the parameters  $q, V_{\rm 01},  V_{\rm 02}$ in such a way that it will be repulsive and attractive at short and long distances, respectively~\cite{calogero1967}, see Section~\ref{method2} for details.
 
In plasma physics the screened Coulomb interaction is often accounted for by means of the  exponential cosine (EC) screened Coulomb  potential $V^{\rm EC} \left(r\right)= V_0^{\rm EC} \exp\left(-qr\right)\cos\left(q r\right)/r$~\cite{lam1972,  shukla2008, shukla2012, lin2010, moldabekov2015, qi2016, munjal2017}. Note that both potentials $V^{\rm EC}$ and $V^{\rm TF}$ derive directly from the random phase approximation (RPA), thus neglecting the short-range exchange and correlation effects in electronic systems~\cite{giuliani2005}.

Finally, our study will  include the square-well (SW) potential $V^{\rm SW}$ that plays a fundamental role in the modeling of short-range interactions in various physical systems. In particular, the SW potential is useful  as first approximation to the nuclear interaction between a neutron and a proton in a deuteron~\cite{krane1988}. The attractive  $V^{\rm SW}$  is defined by~\cite{capri2002}

\begin{equation}\label{eq:squareWell1}
V^{\rm SW}=\left\{
\begin{array}{l}
-V_0,  \quad  r \leq a\\
0,   \;  \qquad     r > a\\
\end{array}
\right.
\end{equation}
where the positive quantities $V_0, a$ defines the SW potential's depth and radius, respectively. For a given collision energy $ k^{2}$, the phase shift  $\delta_0$  is given by~\cite{capri2002}

\begin{equation}\label{eq:swformula1}
\tan \delta_0 = \frac{k j'_0 \left(ka \right) j_0 \left(\alpha a \right) - \alpha j_0 \left(k a \right)j'_0 \left(\alpha a \right) }{k n'_0 \left(ka \right) j_0 \left(\alpha a \right) - \alpha n_0 \left(k a \right)j'_0 \left(\alpha a \right)} \, ,
\end{equation}
where the prime denotes differentiation with respect to the function's argument, and  $\alpha^{2} = k^{2} + V_0 $. Here the symbols  $j_0, n_0$ denote the spherical Bessel functions. Note that Eq.~ \ref{eq:swformula1} is beset by modulo $\pi$ ambiguity, i.e. $ \delta_0 $ is defined up to addition of an arbitrary multiple of  $\pi$ \cite{taylor1972, meijer1975}. 

\subsection{Computation of the Quantum Phase Shifts: The Variable Phase Approach}\label{methodVPM}

The standard computation of the phase shifts takes place in two stages. First, given a spherically symmetric regular potential $V^{\rm sc}\left(r \right)$, one needs to solve  the  radial Schr\"{o}dinger equation 
\begin{equation}\label{eq:schr}
u''_l  \left(r \right) + \left[k^{2} -V^{\rm cf}\left(r \right) - V^{\rm sc}\left(r \right) \right]u_l \left(r \right) = 0 \, ,
\end{equation}
for $u_l\left(r \right)$ with the initial condition  $u_l\left(0\right)=0$ at the origin for a given angular momentum number $l$ ($l=0, 1, \cdots$). The repulsive centrifugal potential $V^{\rm cf}\left(r \right)= l\left(l + 1 \right)/r^{2}$  vanishes in our case as we shall limit ourselves to  the computation of the s-wave phase shift $\delta_0$ only. Therefore Eq.~\ref{eq:schr} in s-wave becomes
\begin{equation}\label{eq:schr1}
u''_0  \left(r \right) + \left[k^{2}  - V^{\rm sc}\left(r \right) \right]u_0 \left(r \right) = 0 \, .
\end{equation}
The solution  $u_0$  can be obtained by numerically integrating Eq.~\ref{eq:schr1}. Note that this task requires the knowledge of the range of $V^{\rm sc}$ in order to avoid possible inaccuracies and round-off errors as well. Finally, the quantity $\delta_0$ is obtained by matching the numerical solution with the free solution of Eq.~\ref{eq:schr1}  in the asymptotic region, i.e., where the effects of $V^{\rm sc}$ are negligible. As a consequence one needs to solve an algebraic equation for $\delta_0$ of the same kind of Eq.~\ref{eq:swformula1}, thereby involving the spherical Bessel functions $j_0, n_0$, and their corresponding derivatives. In general, the above standard approach can be cumbersome to be implemented for in-house computer codes. 

An alternative approach to the computation of the phase shifts is given by the variable phase method~\cite{calogero1963, calogero1967, babikov1967, morse1933}. Despite the fact that the VPM follows from a simple mathematical property, that is, a homogeneous second order differential equation such as the radial  Schr\"{o}dinger equation can be reduced to a first order nonlinear equation of Riccati type, it really presents a paradigmatic shift in the partial wave analysis as the  phase shifts are no longer regarded as mere scalar quantities. In fact, not only the VPM yields the value of the phase shift directly, but  promotes it from a scalar quantity to a phase function $r \mapsto  \delta_0\left(r \right) $, so providing important physical insights about the collision effects due to the presence of a particular potential~\cite{calogero1967}.

The variable phase method rests on the following differential equation for the function $\delta_0\left(r\right)$ 
\begin{equation}\label{eq:phase}
\delta'_0 \left(r \right) = - \frac{V^{\rm sc}\left(r \right) }{k}  \left[\cos\delta_0\left(r \right)	\hat{j_0} \left(kr \right)  - \sin\delta_0\left(r \right)\hat{n}_0\left(kr \right)  \right]^{2}\, , 
\end{equation}
with  the boundary condition at the origin  $\delta_0\left(0\right)=0$. In Eq.~\ref{eq:phase} the symbols $\hat{j_0}$, $\hat{n}_0$ denote the Riccati-Bessel functions. 


The numerical integration of Eq.~\ref{eq:phase} yields the value of $\delta_0$ directly by taking the asymptotic limit of the solution, i.e. $\displaystyle\lim_{r\to \infty} \delta_0\left(r \right) = \delta_0~ $\cite{calogero1967, marchetti2019}. Moreover, the VPM automatically removes the modulo $\pi$ ambiguity~\cite{calogero1967, chadan2001, marchetti2019}. The latter property gives the following expression for the Levinson's theorem~\cite{levinson1949, calogero1967} 
\begin{equation}\label{eq:levinson}
\lim_{k \to 0}   \delta_0\left(k\right) = n_{b, 0}  \pi \, ,
\end{equation}
where $n_{b, 0}$ is the number of bound states that the scattering potential is capable to form in s-wave at low energy. 

\begin{figure}[htp]
\resizebox{0.50\textwidth}{!}{%
  \includegraphics{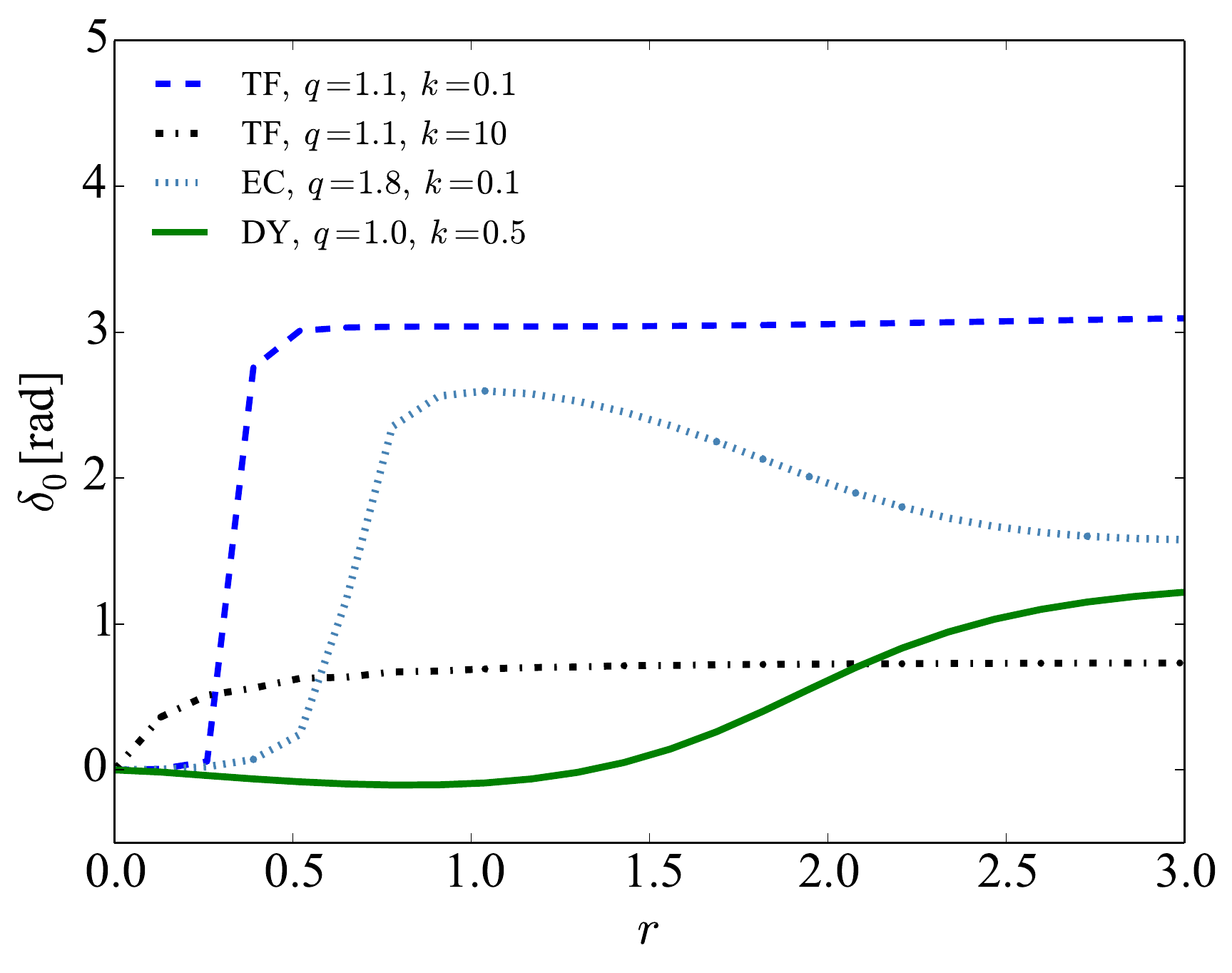}
}
\caption{The phase shift $\delta_0$ as function of $r$ caused by the TF, EC, DY potentials' instances chosen at random from the data sets, computed by means of the VPM for different values of $k$ and $q$ of interest. We refer the reader to  Table~\ref{table:table1} for all the information about the potentials' parameters, and  their respective asymptotic values of  $\delta_0$.}
\label{fig:figure1}
\end{figure}

In Fig.~\ref{fig:figure1} the phase shift $\delta_0$ as function of $r$ due to the presence of an attractive TF potential with $V_{0}^{\rm TF}=5$, $q=1.1$ obtained by means of the VPM for  $k=0.1$ is shown. It is evident that after an abrupt change the phase function  $\delta_0\left(r \right)$ converges to  $\delta_0 \approx \pi$ in the asymptotic region. Therefore, according to Levinson's theorem, see  Eq.~\ref{eq:levinson}, this TF potential's instance supports one bound state in s-wave. In such a case its  respective phase function takes a typical step-like behaviour~\cite{calogero1967, portnoi1997} as displayed in Fig.~\ref{fig:figure1}. However, the information about the presence of one or more bound states can be completely lost by  looking only at the shape of the phase function, when the collision energy is increased.  This fact is illustrated in Fig.~\ref{fig:figure1} by means of the plot of the phase function"s curve corresponding to the above TF potential's instance obtained at  $k=10$. 

Also, in Fig.~\ref{fig:figure1}  the phase function's curves corresponding to the EC and DY potentials with $V_{0}^{\rm EC}=5, q=1.8$  and   $V_{02}= 2V_{01}=20, q_2= 2 q_1 = 2.0 $, computed at  $k=0.1$ and $k=0.5$, respectively, are displayed. All the curves in Fig.~\ref{fig:figure1}, see also Table~\ref{table:table1} for details, correspond to scattering potentials' instances chosen at random from the respective data sets on which the CNN are trained. They clearly exemplify the different learning tasks that the CNN has to undertake in order to approximate $r \mapsto  \delta_0\left(r \right) $ at given collision energy $k^{2}$ for the scattering potential $V^{\rm sc}$ under scrutiny.

\begin{table}
\renewcommand{\arraystretch}{2} 
    \caption{ The strengths and screening parameters of the TF, EC, DY potentials  along with the respective asymptotic phase shifts computed by means of the VPM at different values of $k$ are shown, respectively. Note that $V_{02}= 2V_{01}=20 $ and  $q_2= 2 q_1 = 2.0 $   for the DY potential. }
\label{table:table1}
\begin{ruledtabular}
\begin{tabular}{c c c c   }
$V^{\rm sc}$  & parameters & $k$ &  $\delta_0$  \\ [1ex]
\hline
    $V^{\rm TF}$     & $V_{0}^{\rm TF}=5, q=1.1$       & $0.1 $        & $3.15 $   \\
   $V^{\rm TF}$     & $V_{0}^{\rm TF}=5, q=1.1$         & $10$         & $0.73$    \\
   $V^{\rm EC}$     & $V_{0}^{\rm EC}=5, q=1.8$          & $0.1$         & $1.61$    \\
    $V^{\rm DY}$    & $V_{01}=10, q_1=1.0$           & $0.5$         & $1.28$     \\ [1ex]
\end{tabular}
\end{ruledtabular}
\end{table}

\subsection{Image of the Scattering Potential Instance as Suitable Descriptor} 
\label{method2}

\begin{figure}[h!]
    \hspace*{-0.6cm}\resizebox{0.6\textwidth}{!}{
      \includegraphics{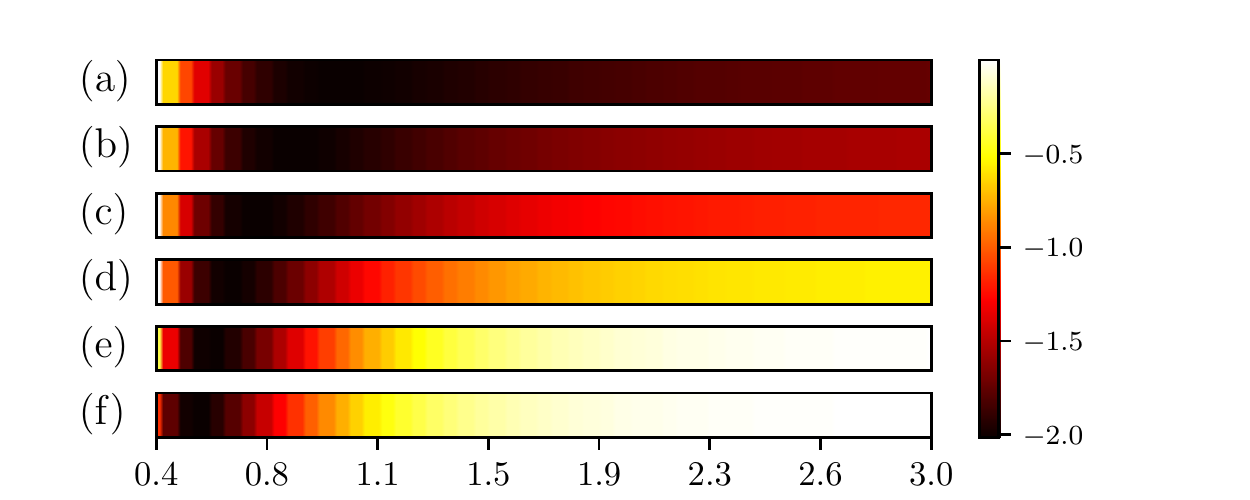}
    }
    \caption{The images $X$ labelled by (a), (b), (c), (d), (e), (f)  relative to the scattering potential $V^{\rm DY}$ with  $V_{01}=10, V_{02}= 20 $ and  $q_1 \equiv q, q_2= 2 q$ obtained for $q = 1.0, 1.2, 1.42, 1.62, 1.8, 1.99$,  respectively. The color (pixel intensity) indicates the value of the potential.}
    \label{fig:plot_images}
    \end{figure}

In our ML model,  the descriptor of  a scattering potential  $V^{\rm sc}$'s instance is  an image $X$ of $N$ pixels, i.e., its dimension $N$ is the resolution of such an image, containing the values of this potential evaluated at $N$ different points within a discretized support interval $I_s = \left[ r_{0}, r_{1}, r_{2} ..., r_{N} \right] $, i.e., $X = \left[ V^{\rm sc}(r_{0}), V^{\rm sc}( r_{1}), V^{\rm sc}( r_{2}), ..., V^{\rm sc}( r_{N}) \right] $.
The minimum  $r_0$ and  the maximum  $r_{N}$ of the set $I_s$ are chosen according to the typical behaviour of the scattering potential $V^{\rm sc}$   in the neighborhood of the origin and its asymptotic regime, respectively. Thus, for a fixed strength of $V^{\rm sc}$ the resolution $N$  would determine a coarse-grained view of such a potential, thereby causing the image's details to vary smoothly with the  screening parameter $q$.  In Fig.~\ref{fig:plot_images} six images $X$ corresponding to different instances of $V^{\rm DY}$ with  parameters $V_{01}=10, V_{02}= 20 $ and  $q_1 \equiv q, q_2= 2 q$  ($q \in [1,2]$) such that it is repulsive at short distance and attractive in the asymptotic region, are shown. Our proposed descriptor offers a good compromise between the accuracy and the computation time.

However, it is not possible to apply such a representation to the potential $V^{\rm SW}$ as its image on the discretized interval $I_s$ 
would not always encode small variations with respect to its screening parameter $1/a$. To overcome this issue we shall resort to the effective range approximation~\cite{krane1988, bethe1949} that holds for very small  values of $k$  only, see Section~\ref{results} for details.

Finally, it is worth noting that the statistical learning  performed by the CNN will be done assuming that each  scattering potential's strength is fixed. The reason is that the phase shift $\delta_0$ caused by an instance of the same potential but with a different strength can be directly derived by an appropriate scaling~\cite{calogero1967}. 

\subsection{Data Sets}\label{datasets}

For each scattering potential, except the SW potential, a data set   $D_k =\{(X_1, \delta_0^{(1)}), (X_2, \delta_0^{(2)}), \cdots ,(X_n, \delta_0^{(n)}) \}$ where $n=1000$ is constructed for $k=0.1, 0.5, 5, 10$, respectively. Each image descriptor $X_{\mu}$ is labelled by the corresponding phase shift $ \delta_0^{(\mu)}$ computed at fixed wave number $k$ by means of the VPM whose implementation rests on the algorithm LSODA~\cite{petzold1983, palov2021}. We divide each data set $D_k$ into training, validation and test sets. The CNN's learning is performed within the training set, while the validation set is used to adjust the model hyperparameters by studying the model convergence during the learning phase. Finally, the assessment of the CNN's accuracy is performed on the test set.

The TF and EC potentials' parameters  upon which the respective data set $D_k$ are constructed are  $V_{0}^{\rm TF}=V_{0}^{\rm EC}=5$ and $q$ being uniformly generated in $ [0.5, 2.5]$ and  $ [0.5, 2]$, respectively. According to this choice of parameters some instances of TF potential can form  s-wave bound states, while each instance of EC potential cannot support them.  In this regard, we estimated that roughly  $30 \%$ instances of TF potentials in $D_{0.1}$ can form one bound state.

The parameters of interest for the DY potential were introduced in Section~\ref{method2}. In the latter case the screening parameter $q_1$ is uniformly chosen at random in  $ [1, 2]$. Note that none of the DY potential's instances  is capable of forming bound states. 

Finally, two data sets  of  shallow SW potentials with depth  $V_0=0.25$ and radii $a$ uniformly chosen in  $[0.4, 2]$ are generated computing the respective s-wave phase shifts  according to Eq.~\ref{eq:swformula1} at $k=0.01, 0.1$.

\subsection{Convolutional Neural Network} \label{neural}

Our model is composed of a linear stack of $L=5$ one-dimensional convolutional layers each followed by a max-pooling layer. In each convolutional layer different numbers of filters, each of size $3$ pixels, take the output of the previous layer (or the input image in case of the first layer) and extract features while scanning one pixel at time. Each convolution is followed by the Rectified Linear Unit (ReLU) activation. In this way, the computation units of a convolutional layer create feature maps that summarise the presence of relevant details in the input images.

After each convolution layer, a max-pooling layer encapsulates the most activated output of the feature map, resulting in a down sampled representation of the layer input. As the information is processed through the layers of the network, convolutional and max-pooling operations make the model able to capture patterns of progressively higher complexity. Hence the number of filters in each layer are progressively increased to allow  more abstraction. In particular, we used $64, 128, 256, 256, 256$ filters on the first to the last convolutional layer, respectively. At the end of the network a series or two fully connected layers, the first one with output size $1024$, and the last one with a single output, perform label prediction based on the features extracted by the last max-pooling layer.
The CNN have been implemented on TensorFlow~\cite{tensorflow} (version 2.4.0), and trained by means of the Adam optimizer~\cite{adam_opt}.


\section{Results and Discussion} \label{results}

    \begin{figure}
    \hspace*{-0.1cm}\resizebox{0.5\textwidth}{!}{%
      \includegraphics{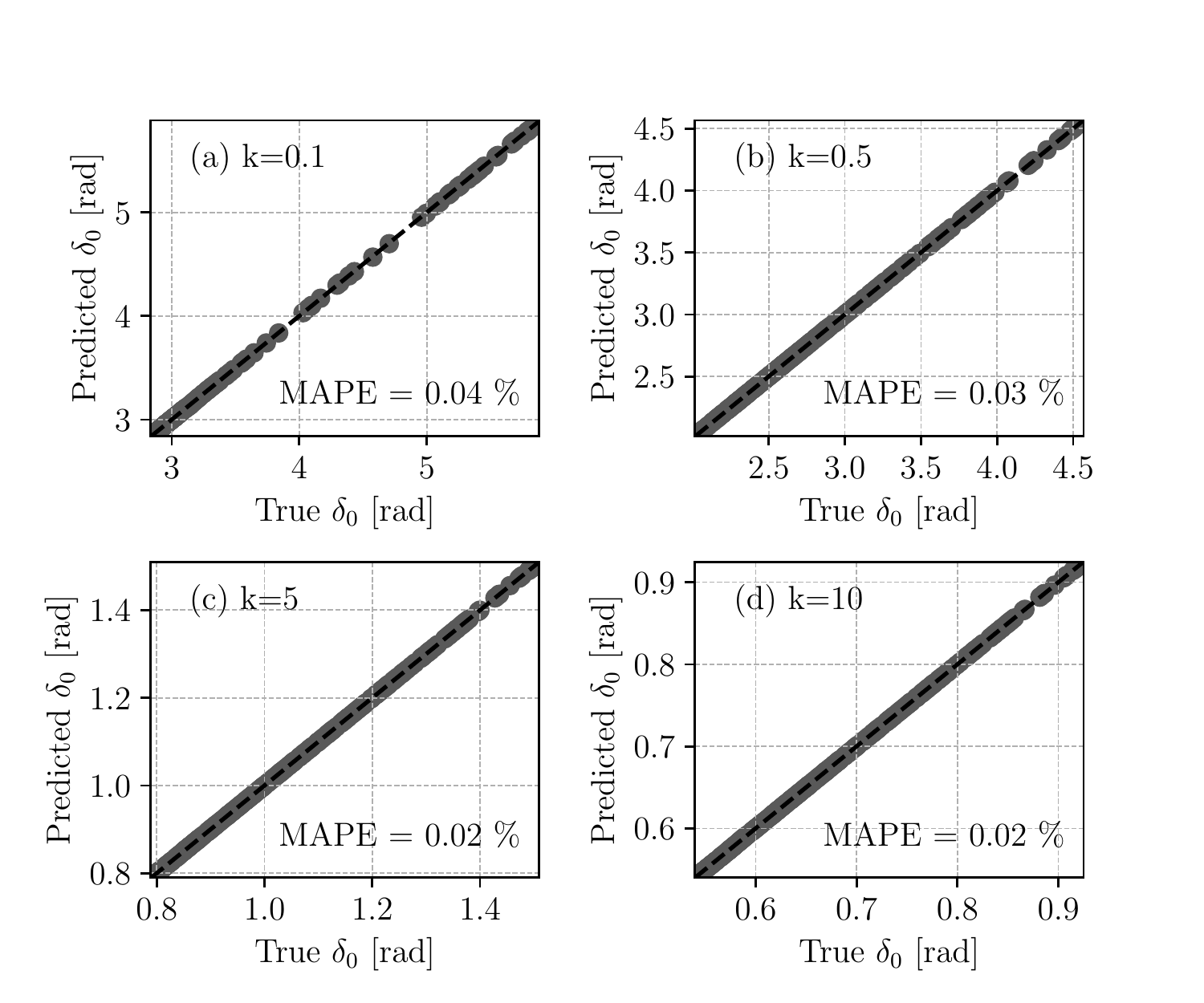}
    }
    \hspace*{-0.1cm}\resizebox{0.5\textwidth}{!}{%
      \includegraphics{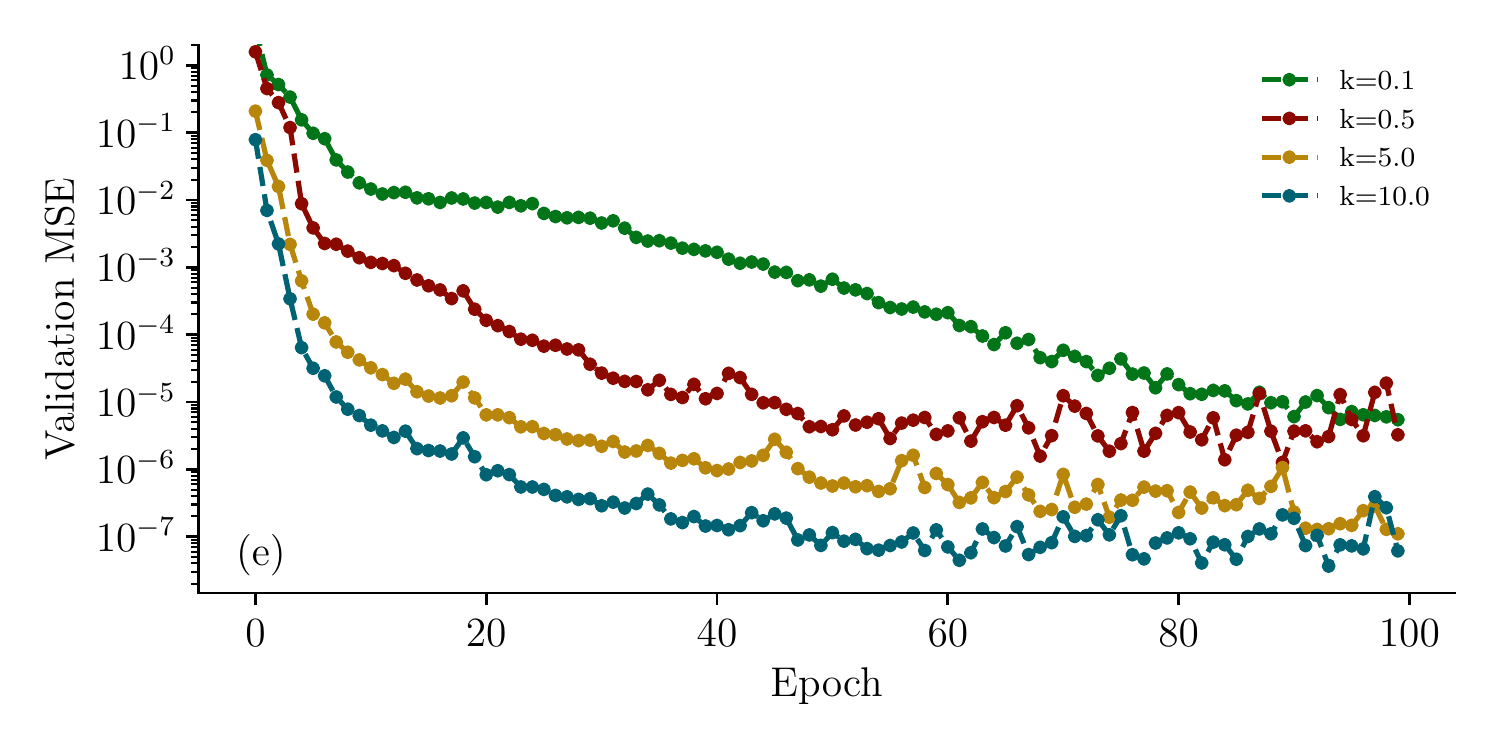}
    }
    
    \caption{In panels (a), (b), (c), (d) predictions versus reference phase shifts $\delta_0$ caused by the TF potential at  $k=0.1, 0.5, 5, 10$, respectively, are shown. Panel (e) displays the relative learning curves as functions of the training epochs.}
    \label{fig:scatter_tf}
    \end{figure}
    
    \begin{figure}[!t]
    \hspace*{-0.2cm}\resizebox{0.5\textwidth}{!}{%
      \includegraphics{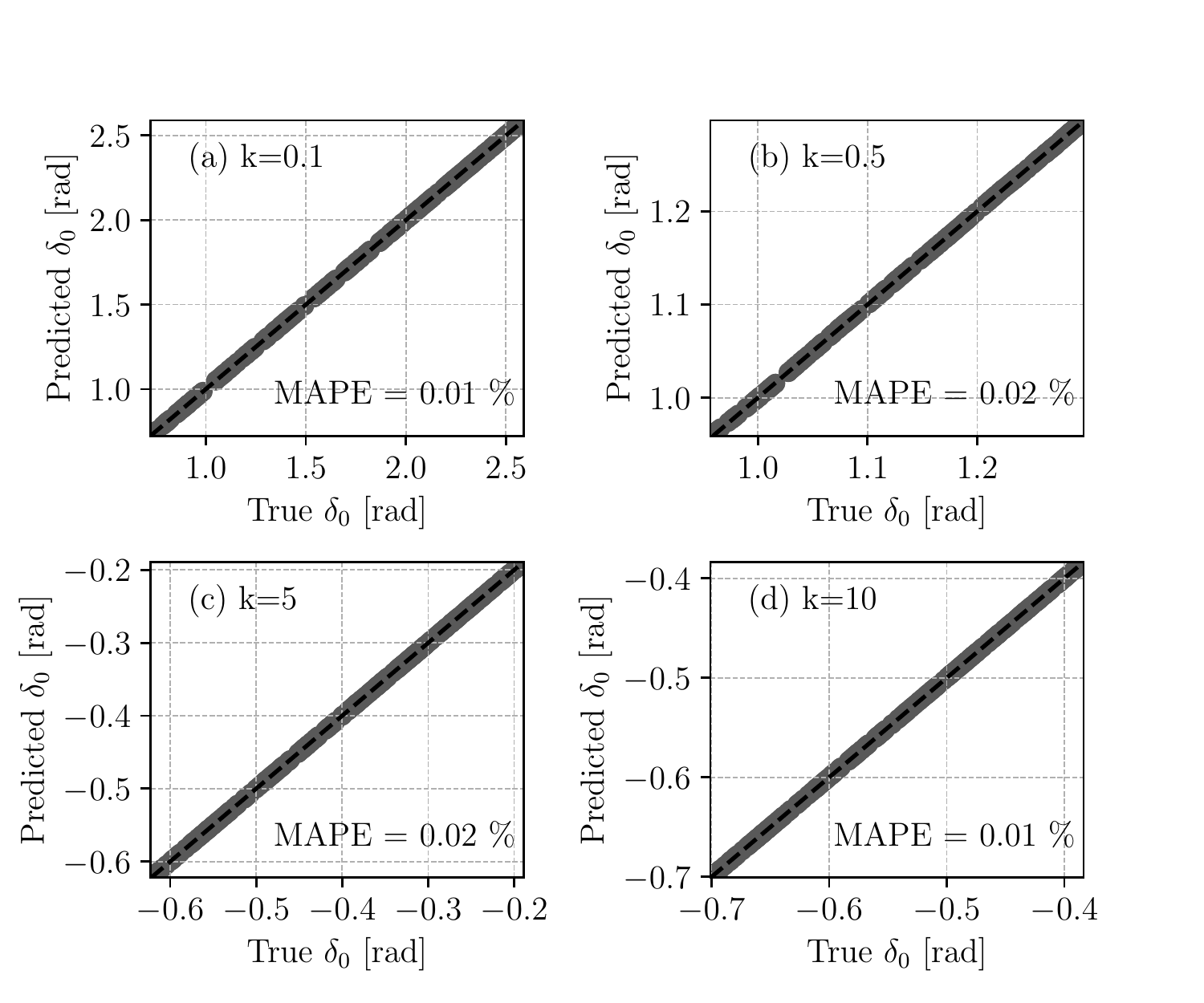}
    }
    \hspace*{-0.1cm}\resizebox{0.5\textwidth}{!}{%
      \includegraphics{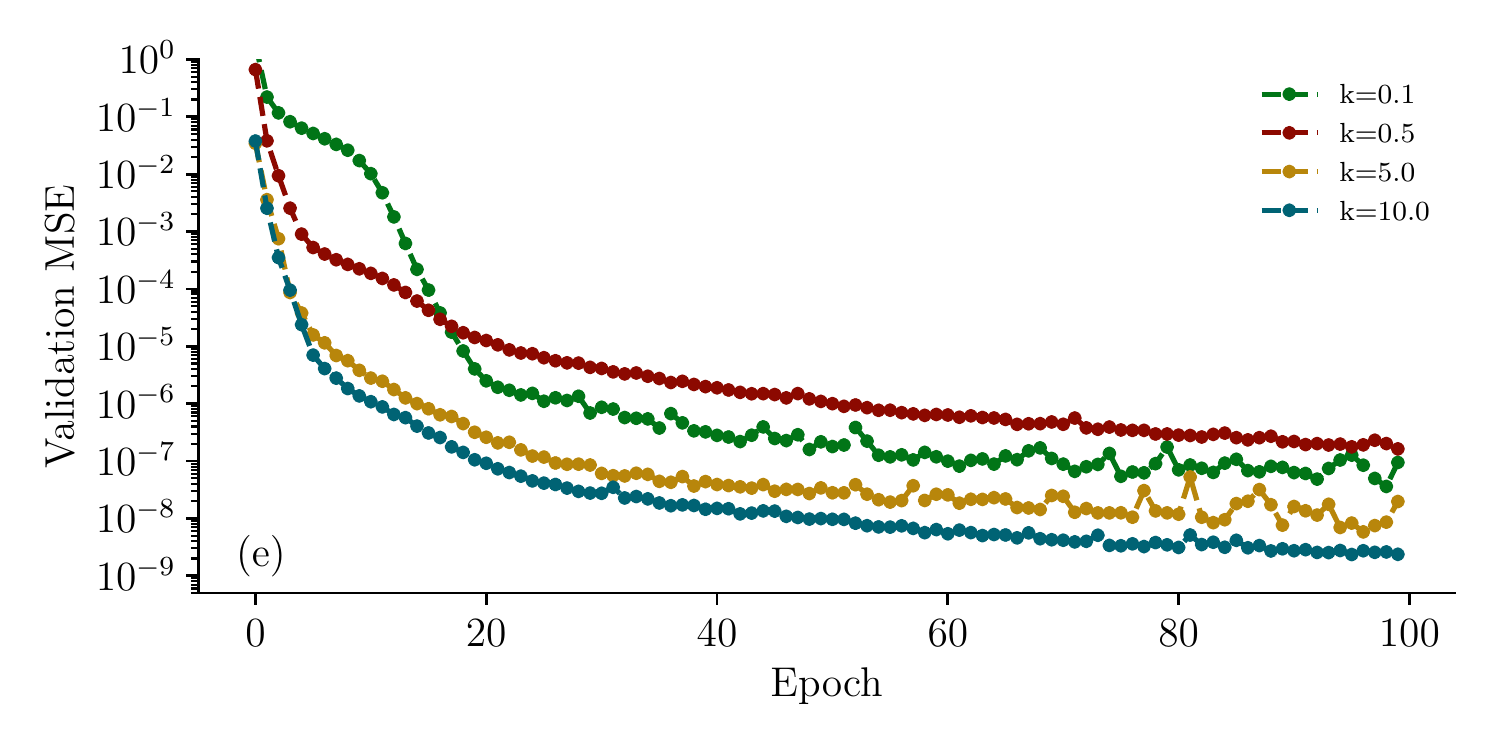}
    }

    \caption{In panels (a), (b), (c), (d) predictions versus reference phase shifts $\delta_0$ caused by the DY potential at  $k=0.1, 0.5, 5, 10$, respectively, are shown. Panel (e) displays the relative learning curves as functions of the training epochs.}
    \label{fig:double_yukawa_scatter}
    \end{figure}

    \begin{figure}
    \hspace*{-0.2cm}\resizebox{0.5\textwidth}{!}{%
      \includegraphics{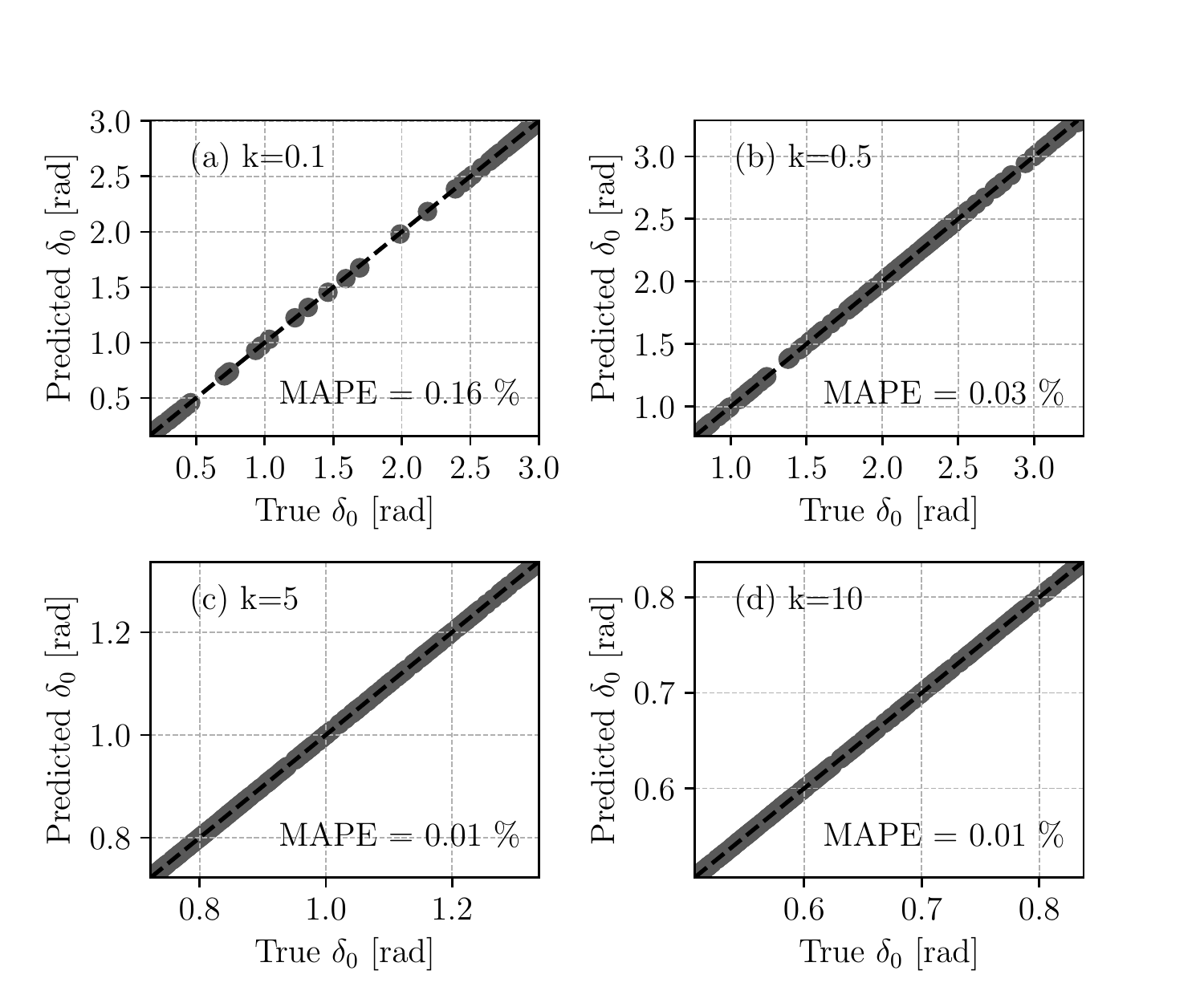}
    }
    \hspace*{-0.1cm}\resizebox{0.5\textwidth}{!}{%
      \includegraphics{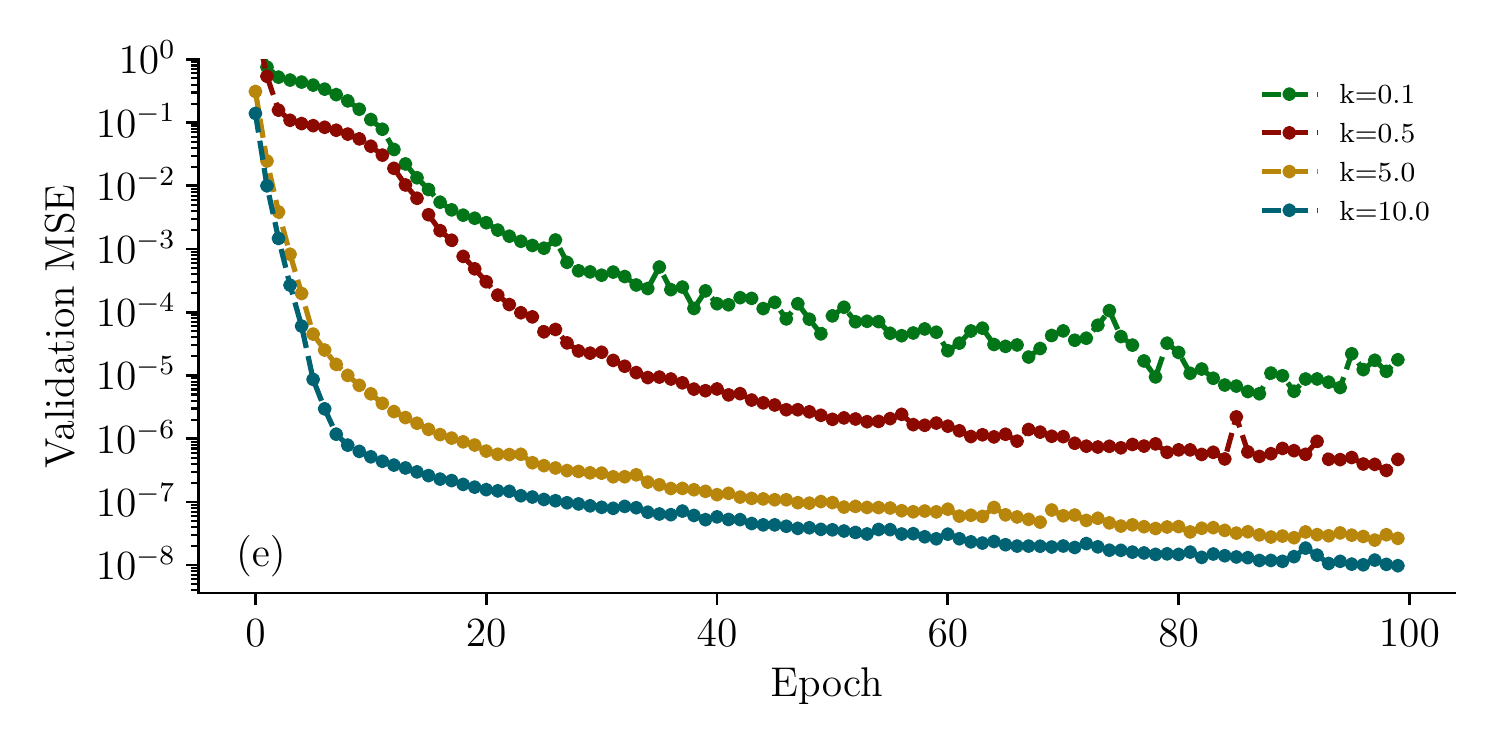}
    }

    \caption{In panels (a), (b), (c), (d) predictions versus reference phase shifts $\delta_0$ caused by the EC potential at  $k=0.1, 0.5, 5, 10$, respectively, are shown. Panel (e) displays the relative learning curves as functions of the training epochs. }
    \label{fig:scatter_ec}
    \end{figure}
    
In order to assess  the CNN's performance over the above data sets we adopt two estimator measures: the mean square error (MSE) and  mean absolute percentage error. First, we train the CNN by minimizing the MSE during the training phase using the Adam optimizer. Second, we report the MAPE between the true and predicted phase shifts in the test set as a measure of model performance. Note that unless differently specified, the training happened over 100 epochs on 16 mini batch using a learning rate of $1 \times 10^{-4}$.

In panels (a), (b), (c), (d) of Fig.~\ref{fig:scatter_tf}, Fig.~\ref{fig:double_yukawa_scatter} and Fig.~\ref{fig:scatter_ec} the scatter plots of the prediction versus the reference s-wave phase shift caused by TF, DY, and EC, respectively,  at  $k=0.1, 0.5, 5, 10$ are shown. For all these potentials the MAPE  is comparable and varies from $0.01\%$ to $0.03\%$ when the learning occurs for  $ k=0.5, 5, 10$. There is a  plausible physical explanation of these similar learning performance obtained despite the different functional form of the scattering potentials under study. The reason is that increasing the energy for the screening parameters under scrutiny each scattering potential affects the phase function in the neighbourhood of the origin only. To understand this, one just needs to look at the scatter plot relative to DY potential in panels (c) and (d) of Fig.~\ref{fig:double_yukawa_scatter} where the phase shifts are negative. This happens because the repulsive part of the DY potentials' instances  mainly  affect the phase function nearby the origin as a consequence to our choice of the screening parameters. On the other hand, by inspections of some  instances of the  TF, DY, and EC potentials' data sets  we found that for the parameters of interest, increasing the energy the phase function  $r \mapsto  \delta_0\left(r \right) $ takes a smooth form for each $V^{\rm sc}$. For instance,  this smooth behaviour is illustrated by the plot of the phase function due to a TF potential's instance at $k=10$ in Fig.~\ref{fig:figure1}. Thus, the CNN's learning task become easier at higher collision energies, that is, increasing $k$. 
On the other hand this observation is confirmed looking at the learning curves as functions of epochs shown  in panel (e) of Fig.~\ref{fig:scatter_tf}, Fig.~\ref{fig:double_yukawa_scatter} and Fig.~\ref{fig:scatter_ec} for $k=5, 10$ where a far smaller number of training epochs with respect to that observed for $k=0.1, 0.5$ is necessary for achieving the convergence. As a consequence the MAPE increases at low energies.

As the Adam optimizer algorithm takes considerably more epochs to converge over the data set corresponding to small values of $k$, it is evident that the CNN's learning task becomes difficult.  This can be easily understood looking at the data set  $D_{0.1}$ for the Thomas-Fermi potential $V^{\rm TF}$. In such a case we know that it contains about $30 \%$ of  $V^{\rm TF}$ instances supporting one bound state. Therefore in such a case the CNN  need to approximate a phase function $\delta_0\left(t\right)$ presenting an abrupt change due to the emergence of a plateau where $\delta_0\approx \pi$ for certain values of the screening parameter $q$. We illustrated this characteristic behaviour due to presence of the bound states in Section~\ref{methodVPM} (see also Fig.~\ref{fig:figure1}). Moreover,  the phase functions relative to $V^{\rm EC}$ and $V^{\rm DY}$ certainly exhibit a far richer functional form at low energy as illustrated by two respective phase functions chosen at random  in Fig.~\ref{fig:figure1}.  In particular, some $V^{\rm EC}$ instances can be repulsive, attractive and repulsive, respectively, in three different contiguous regions, respectively. This behaviour increases the  functional complexity of the phase function, thus  explaining the reason for which the largest MAPE  $\approx 0.16$ is obtained for the EC potential at $k=0.1$ (see  the relative scatter plot in panel (a) of Fig.~\ref{fig:scatter_ec}).

 Overall, the previous observations explain why the CNN's   performance in general worsens slightly at small values of $k$. Despite this fact,  our ML model yields accurate predictions of the phase shifts caused by TF, EC, DY potentials with relative error that is much less than $1\%$. 

As explained in Section~\ref{method2} it is not possible to meaningfully represent the SW potential as an image  in our approach. To overcome this issue, we can resort the effective range approximation for short-range potentials \cite{krane1988, bethe1949} that is applicable in s-wave scattering at low energy, i.e., $k  \approx 0$. This approximation reads as~\cite{bethe1949}
\begin{equation}\label{eq:bethe}
k \cot \delta_0 \approx -\frac{1}{c_0} + \frac{1}{2} k^{2} r_{0} \, ,
\end{equation}
where $c_0, r_0$ are called the scattering length and the effective range, respectively. In general, for simple potentials such as the SW and TF potentials, $\delta_0$ depends only on the depth and range of the scattering potential under study \cite{capri2002}. From Eq.~\ref{eq:bethe}  it is then possible to fix the TF potential's parameters $ V_{0}^{\rm TF}, q$ in terms of $V_0, a$ relative to the SW potential, so that the former yields the same s-wave shifts $\delta_0$ caused by the SW potential. Note that this approach is often called the shape independent approximation (SIA). To implement the SIA, one needs to compute $c_0, r_0$ for both potentials. By their comparison, one finds the following  relations $ V_{0}^{\rm TF} = \left(1/3\right)
\left(5/2 \right)^{3/2} V_0, q = \left(5/2 \right)^{1/2}a^{-1}$. Next,  
we can represent a given  instance of a SW potential, labelled by $\delta_0$ according to  Eq.~\ref{eq:swformula1},   by the image of a TF potential whose parameters satisfy the above SIA's relations.

\begin{figure}
  \hspace*{-0.2cm}\resizebox{0.5\textwidth}{!}{%
    \includegraphics{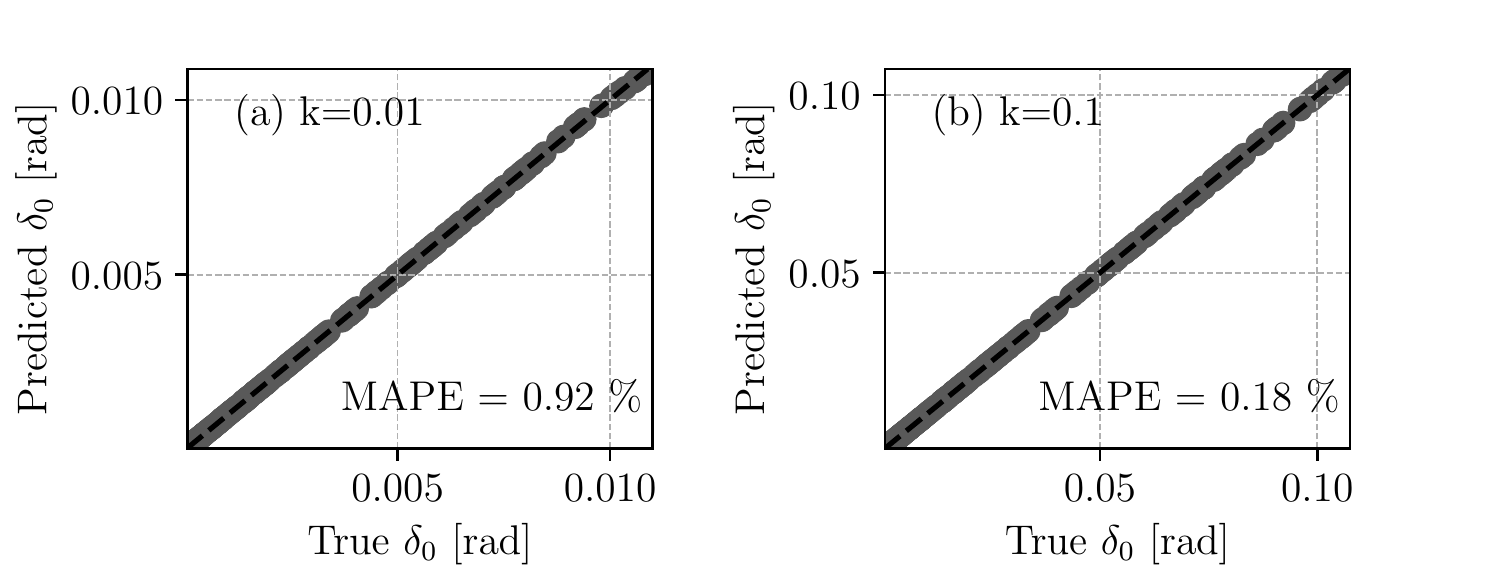}
  }
  \caption{In panels (a), (b) predictions versus reference phase shifts $\delta_0$ caused by the SW potential replaced by the TF potential according to the shape independent approximation at  $k=0.01, 0.1$, respectively, are shown.}
  \label{fig:scatter_sqw}
  \end{figure}

In panels (a), (b) of Fig.~\ref{fig:scatter_sqw} the scatter plots of the prediction versus the reference s-wave phase shift according to the SIA applied to the SW potential~\footnote{Note that in this case, using the same model hyperparameters as in the previous cases, resulted in higher variability in validation loss during training. We regularized the model by reducing the learning rate of the Adam optimizer to $10^{-6}$ and increasing the training epochs increased to 5000.}, at  $k=0.01, 0.1$, respectively, are shown. We found MAPE is  $0.92\%$, $0.18\%$, for $k=0.01, 0.1$ respectively. Therefore  this scattering approximation combined with our ML model can be used to yield satisfactory accurate predictions of $\delta_0$ even for nonrelativistic collisions in the presence of a square-well potential.

Our analysis suggests that our ML model is effectively learning the phase functions after being trained on the scattering potential images  $X_{\mu}$  labelled by the respective phase shift $ \delta_0^{(\mu)}$.

\section{Summary and Outlook} \label{conclusion}

We presented a CNN model  for learning the s-wave phase shifts arising from four different three-dimensional spherically symmetric potentials. The good machine learning performance achieved by our model demonstrates  that the Hamiltonian can serve as a guiding principle in the construction of a physically-motivated descriptor. 

Moreover, we tried to understand how the CNN's learning tasks vary with the functional complexity of the phase function emerging in the presence of different potentials. To this aim, we  add a number of instances of the Thomas-Fermi potential supporting bound states to the respective data set. In such a case, we found that the CNN's regression task becomes difficult requiring a substantial increase of the number of epochs for reaching a good performance. We argue that this larger number of epochs mainly arises from the steepness of the mapping  $r \mapsto  \delta_0\left(r \right) $ as a direct consequence of the Levinson's theorem, thereby hindering the CNN's learning task.

The next goal could be to build a supervised machine learning model able to predict $\delta_l$ with $l=1, 2, \cdots$ at fixed collision energy by including the centrifugal potential contribution into the descriptor in a similar way that we discussed here. However, a more ambitious goal would be to effectively learning $r \mapsto  \delta_l\left(r,k \right) $ having the wave number $k$ as additional input in the model.

\begin{acknowledgments}
We thank professor Ravi Rau for pointing out important works regarding the phase amplitude equation and dr. Jan Hermann and dr. Peter Šušnjar for valuable comments and suggestions. This work was supported by the EU through the European Regional Development Fund CoE program TK133 “The Dark Side of the Universe”.
\end{acknowledgments}

\section*{Author Contribution Statement}
Both authors contributed equally to this study.

\section*{Data Availability}

The data that support the findings of this study are available from the corresponding author upon reasonable request.

\appendix

\bibliography{references}{}

\begin{thebibliography}{75}%
\makeatletter
\providecommand \@ifxundefined [1]{%
 \@ifx{#1\undefined}
}%
\providecommand \@ifnum [1]{%
 \ifnum #1\expandafter \@firstoftwo
 \else \expandafter \@secondoftwo
 \fi
}%
\providecommand \@ifx [1]{%
 \ifx #1\expandafter \@firstoftwo
 \else \expandafter \@secondoftwo
 \fi
}%
\providecommand \natexlab [1]{#1}%
\providecommand \enquote  [1]{``#1''}%
\providecommand \bibnamefont  [1]{#1}%
\providecommand \bibfnamefont [1]{#1}%
\providecommand \citenamefont [1]{#1}%
\providecommand \href@noop [0]{\@secondoftwo}%
\providecommand \href [0]{\begingroup \@sanitize@url \@href}%
\providecommand \@href[1]{\@@startlink{#1}\@@href}%
\providecommand \@@href[1]{\endgroup#1\@@endlink}%
\providecommand \@sanitize@url [0]{\catcode `\\12\catcode `\$12\catcode
  `\&12\catcode `\#12\catcode `\^12\catcode `\_12\catcode `\%12\relax}%
\providecommand \@@startlink[1]{}%
\providecommand \@@endlink[0]{}%
\providecommand \url  [0]{\begingroup\@sanitize@url \@url }%
\providecommand \@url [1]{\endgroup\@href {#1}{\urlprefix }}%
\providecommand \urlprefix  [0]{URL }%
\providecommand \Eprint [0]{\href }%
\providecommand \doibase [0]{http://dx.doi.org/}%
\providecommand \selectlanguage [0]{\@gobble}%
\providecommand \bibinfo  [0]{\@secondoftwo}%
\providecommand \bibfield  [0]{\@secondoftwo}%
\providecommand \translation [1]{[#1]}%
\providecommand \BibitemOpen [0]{}%
\providecommand \bibitemStop [0]{}%
\providecommand \bibitemNoStop [0]{.\EOS\space}%
\providecommand \EOS [0]{\spacefactor3000\relax}%
\providecommand \BibitemShut  [1]{\csname bibitem#1\endcsname}%
\let\auto@bib@innerbib\@empty
\bibitem [{\citenamefont {Joachain}(1975)}]{joachain1975}%
  \BibitemOpen
  \bibfield  {author} {\bibinfo {author} {\bibfnamefont {C.~J.}\ \bibnamefont
  {Joachain}},\ }\href@noop {} {\emph {\bibinfo {title} {{Quantum Collision
  Theory}}}}\ (\bibinfo  {publisher} {North-Holland},\ \bibinfo {year}
  {1975})\BibitemShut {NoStop}%
\bibitem [{\citenamefont {Piel}\ and\ \citenamefont
  {Chrysos}(2020)}]{piel2020}%
  \BibitemOpen
  \bibfield  {author} {\bibinfo {author} {\bibfnamefont {H.}~\bibnamefont
  {Piel}}\ and\ \bibinfo {author} {\bibfnamefont {M.}~\bibnamefont {Chrysos}},\
  }\href@noop {} {\bibfield  {journal} {\bibinfo  {journal} {Molecular
  Physics}\ }\textbf {\bibinfo {volume} {118}},\ \bibinfo {pages} {e1587024}
  (\bibinfo {year} {2020})}\BibitemShut {NoStop}%
\bibitem [{\citenamefont {Fermi}(1936)}]{fermi1936}%
  \BibitemOpen
  \bibfield  {author} {\bibinfo {author} {\bibfnamefont {E.}~\bibnamefont
  {Fermi}},\ }\href@noop {} {\bibfield  {journal} {\bibinfo  {journal} {Ric.
  Scientifica}\ }\textbf {\bibinfo {volume} {7}} (\bibinfo {year}
  {1936})}\BibitemShut {NoStop}%
\bibitem [{\citenamefont {Huang}\ and\ \citenamefont {Yang}(1957)}]{huang1957}%
  \BibitemOpen
  \bibfield  {author} {\bibinfo {author} {\bibfnamefont {K.}~\bibnamefont
  {Huang}}\ and\ \bibinfo {author} {\bibfnamefont {C.~N.}\ \bibnamefont
  {Yang}},\ }\href@noop {} {\bibfield  {journal} {\bibinfo  {journal} {Phys.
  Rev.}\ }\textbf {\bibinfo {volume} {105}},\ \bibinfo {pages} {767} (\bibinfo
  {year} {1957})}\BibitemShut {NoStop}%
\bibitem [{\citenamefont {Idziaszek}\ and\ \citenamefont
  {Calarco}(2006)}]{idziaszek2006}%
  \BibitemOpen
  \bibfield  {author} {\bibinfo {author} {\bibfnamefont {Z.}~\bibnamefont
  {Idziaszek}}\ and\ \bibinfo {author} {\bibfnamefont {T.}~\bibnamefont
  {Calarco}},\ }\href@noop {} {\bibfield  {journal} {\bibinfo  {journal} {Phys.
  Rev. Lett.}\ }\textbf {\bibinfo {volume} {96}},\ \bibinfo {pages} {013201}
  (\bibinfo {year} {2006})}\BibitemShut {NoStop}%
\bibitem [{\citenamefont {Pain}(2018)}]{pain2018}%
  \BibitemOpen
  \bibfield  {author} {\bibinfo {author} {\bibfnamefont {J.-C.}\ \bibnamefont
  {Pain}},\ }\href@noop {} {\bibfield  {journal} {\bibinfo  {journal} {Journal
  of Physics Communications}\ }\textbf {\bibinfo {volume} {2}},\ \bibinfo
  {pages} {025015} (\bibinfo {year} {2018})}\BibitemShut {NoStop}%
\bibitem [{\citenamefont {Kohn}(1948)}]{kohn1948}%
  \BibitemOpen
  \bibfield  {author} {\bibinfo {author} {\bibfnamefont {W.}~\bibnamefont
  {Kohn}},\ }\href@noop {} {\bibfield  {journal} {\bibinfo  {journal} {Phys.
  Rev.}\ }\textbf {\bibinfo {volume} {74}},\ \bibinfo {pages} {1763} (\bibinfo
  {year} {1948})}\BibitemShut {NoStop}%
\bibitem [{\citenamefont {Nesbet}(1968)}]{nesbet1968}%
  \BibitemOpen
  \bibfield  {author} {\bibinfo {author} {\bibfnamefont {R.~K.}\ \bibnamefont
  {Nesbet}},\ }\href@noop {} {\bibfield  {journal} {\bibinfo  {journal} {Phys.
  Rev.}\ }\textbf {\bibinfo {volume} {175}},\ \bibinfo {pages} {134} (\bibinfo
  {year} {1968})}\BibitemShut {NoStop}%
\bibitem [{\citenamefont {Calogero}(1963)}]{calogero1963}%
  \BibitemOpen
  \bibfield  {author} {\bibinfo {author} {\bibfnamefont {F.}~\bibnamefont
  {Calogero}},\ }\href@noop {} {\bibfield  {journal} {\bibinfo  {journal} {Il
  Nuovo Cimento (1955-1965)}\ }\textbf {\bibinfo {volume} {27}},\ \bibinfo
  {pages} {261} (\bibinfo {year} {1963})}\BibitemShut {NoStop}%
\bibitem [{\citenamefont {Calogero}(1967)}]{calogero1967}%
  \BibitemOpen
  \bibfield  {author} {\bibinfo {author} {\bibfnamefont {F.}~\bibnamefont
  {Calogero}},\ }\href@noop {} {\emph {\bibinfo {title} {{ Variable Phase
  Approach to Potential Scattering }}}}\ (\bibinfo  {publisher} {Academic
  Press},\ \bibinfo {year} {1967})\BibitemShut {NoStop}%
\bibitem [{\citenamefont {Babikov}(1967)}]{babikov1967}%
  \BibitemOpen
  \bibfield  {author} {\bibinfo {author} {\bibfnamefont {V.~V.}\ \bibnamefont
  {Babikov}},\ }\href@noop {} {\bibfield  {journal} {\bibinfo  {journal}
  {Soviet Physics Uspekhi}\ }\textbf {\bibinfo {volume} {10}},\ \bibinfo
  {pages} {271} (\bibinfo {year} {1967})}\BibitemShut {NoStop}%
\bibitem [{\citenamefont {Fano}\ and\ \citenamefont {Rau}(1986)}]{fano1986}%
  \BibitemOpen
  \bibfield  {author} {\bibinfo {author} {\bibfnamefont {U.}~\bibnamefont
  {Fano}}\ and\ \bibinfo {author} {\bibfnamefont {A.~R.~P.}\ \bibnamefont
  {Rau}},\ }\href@noop {} {\emph {\bibinfo {title} {{Atomic Collisions and
  Spectra}}}}\ (\bibinfo  {publisher} {Academic Press, Inc.},\ \bibinfo
  {address} {Orlando, Florida},\ \bibinfo {year} {1986})\BibitemShut {NoStop}%
\bibitem [{\citenamefont {Palov}\ and\ \citenamefont
  {Balint-Kurti}(2021)}]{palov2021}%
  \BibitemOpen
  \bibfield  {author} {\bibinfo {author} {\bibfnamefont {A.}~\bibnamefont
  {Palov}}\ and\ \bibinfo {author} {\bibfnamefont {G.}~\bibnamefont
  {Balint-Kurti}},\ }\href {\doibase https://doi.org/10.1016/j.cpc.2021.107895}
  {\bibfield  {journal} {\bibinfo  {journal} {Computer Physics Communications}\
  }\textbf {\bibinfo {volume} {263}},\ \bibinfo {pages} {107895} (\bibinfo
  {year} {2021})}\BibitemShut {NoStop}%
\bibitem [{\citenamefont {Zdeborová}(2016)}]{zdeborova2017}%
  \BibitemOpen
  \bibfield  {author} {\bibinfo {author} {\bibfnamefont {L.}~\bibnamefont
  {Zdeborová}},\ }\href@noop {} {\bibfield  {journal} {\bibinfo  {journal}
  {Nature Physics}\ }\textbf {\bibinfo {volume} {13}},\ \bibinfo {pages} {420}
  (\bibinfo {year} {2016})}\BibitemShut {NoStop}%
\bibitem [{\citenamefont {Carleo}\ \emph {et~al.}(2019)\citenamefont {Carleo},
  \citenamefont {Cirac}, \citenamefont {Cranmer}, \citenamefont {Daudet},
  \citenamefont {Schuld}, \citenamefont {Tishby}, \citenamefont
  {Vogt-Maranto},\ and\ \citenamefont {Zdeborov\'a}}]{carleo2019}%
  \BibitemOpen
  \bibfield  {author} {\bibinfo {author} {\bibfnamefont {G.}~\bibnamefont
  {Carleo}}, \bibinfo {author} {\bibfnamefont {I.}~\bibnamefont {Cirac}},
  \bibinfo {author} {\bibfnamefont {K.}~\bibnamefont {Cranmer}}, \bibinfo
  {author} {\bibfnamefont {L.}~\bibnamefont {Daudet}}, \bibinfo {author}
  {\bibfnamefont {M.}~\bibnamefont {Schuld}}, \bibinfo {author} {\bibfnamefont
  {N.}~\bibnamefont {Tishby}}, \bibinfo {author} {\bibfnamefont
  {L.}~\bibnamefont {Vogt-Maranto}}, \ and\ \bibinfo {author} {\bibfnamefont
  {L.}~\bibnamefont {Zdeborov\'a}},\ }\href@noop {} {\bibfield  {journal}
  {\bibinfo  {journal} {Rev. Mod. Phys.}\ }\textbf {\bibinfo {volume} {91}},\
  \bibinfo {pages} {045002} (\bibinfo {year} {2019})}\BibitemShut {NoStop}%
\bibitem [{\citenamefont {Schmidt}\ \emph {et~al.}(2016)\citenamefont
  {Schmidt}, \citenamefont {Marques},\ and\ \citenamefont
  {Botti}}]{schmidt2019}%
  \BibitemOpen
  \bibfield  {author} {\bibinfo {author} {\bibfnamefont {J.}~\bibnamefont
  {Schmidt}}, \bibinfo {author} {\bibfnamefont {M.~R.~G.}\ \bibnamefont
  {Marques}}, \ and\ \bibinfo {author} {\bibfnamefont {M.~A.~L.}\ \bibnamefont
  {Botti}, \bibfnamefont {Silvana.~Marques}},\ }\href@noop {} {\bibfield
  {journal} {\bibinfo  {journal} {npj Computational Materials}\ }\textbf
  {\bibinfo {volume} {5}},\ \bibinfo {pages} {489} (\bibinfo {year}
  {2016})}\BibitemShut {NoStop}%
\bibitem [{\citenamefont {Faber}\ \emph {et~al.}(2017)\citenamefont {Faber},
  \citenamefont {Hutchison}, \citenamefont {Huang}, \citenamefont {Gilmer},
  \citenamefont {Schoenholz}, \citenamefont {Dahl}, \citenamefont {Vinyals},
  \citenamefont {Kearnes}, \citenamefont {Riley},\ and\ \citenamefont {von
  Lilienfeld}}]{faber2017}%
  \BibitemOpen
  \bibfield  {author} {\bibinfo {author} {\bibfnamefont {F.~A.}\ \bibnamefont
  {Faber}}, \bibinfo {author} {\bibfnamefont {L.}~\bibnamefont {Hutchison}},
  \bibinfo {author} {\bibfnamefont {B.}~\bibnamefont {Huang}}, \bibinfo
  {author} {\bibfnamefont {J.}~\bibnamefont {Gilmer}}, \bibinfo {author}
  {\bibfnamefont {S.~S.}\ \bibnamefont {Schoenholz}}, \bibinfo {author}
  {\bibfnamefont {G.~E.}\ \bibnamefont {Dahl}}, \bibinfo {author}
  {\bibfnamefont {O.}~\bibnamefont {Vinyals}}, \bibinfo {author} {\bibfnamefont
  {S.}~\bibnamefont {Kearnes}}, \bibinfo {author} {\bibfnamefont {P.~F.}\
  \bibnamefont {Riley}}, \ and\ \bibinfo {author} {\bibfnamefont {O.~A.}\
  \bibnamefont {von Lilienfeld}},\ }\href@noop {} {\bibfield  {journal}
  {\bibinfo  {journal} {Journal of Chemical Theory and Computation}\ }\textbf
  {\bibinfo {volume} {13}},\ \bibinfo {pages} {5255–5264} (\bibinfo {year}
  {2017})}\BibitemShut {NoStop}%
\bibitem [{\citenamefont {Bartók}\ \emph {et~al.}(2017)\citenamefont
  {Bartók}, \citenamefont {De}, \citenamefont {Poelking}, \citenamefont
  {Bernstein}, \citenamefont {Kermode}, \citenamefont {Csányi},\ and\
  \citenamefont {Ceriotti}}]{bartok2017}%
  \BibitemOpen
  \bibfield  {author} {\bibinfo {author} {\bibfnamefont {A.~P.}\ \bibnamefont
  {Bartók}}, \bibinfo {author} {\bibfnamefont {S.}~\bibnamefont {De}},
  \bibinfo {author} {\bibfnamefont {C.}~\bibnamefont {Poelking}}, \bibinfo
  {author} {\bibfnamefont {N.}~\bibnamefont {Bernstein}}, \bibinfo {author}
  {\bibfnamefont {J.~R.}\ \bibnamefont {Kermode}}, \bibinfo {author}
  {\bibfnamefont {G.}~\bibnamefont {Csányi}}, \ and\ \bibinfo {author}
  {\bibfnamefont {M.}~\bibnamefont {Ceriotti}},\ }\href@noop {} {\bibfield
  {journal} {\bibinfo  {journal} {Sci Adv.}\ }\textbf {\bibinfo {volume} {3}},\
  \bibinfo {pages} {e1701816} (\bibinfo {year} {2017})}\BibitemShut {NoStop}%
\bibitem [{\citenamefont {Montavon}\ \emph {et~al.}(2013)\citenamefont
  {Montavon}, \citenamefont {Rupp}, \citenamefont {Gobre}, \citenamefont
  {Vazquez-Mayagoitia}, \citenamefont {Hansen}, \citenamefont {Tkatchenko},
  \citenamefont {Müller},\ and\ \citenamefont {von
  Lilienfeld}}]{montavon2013}%
  \BibitemOpen
  \bibfield  {author} {\bibinfo {author} {\bibfnamefont {G.}~\bibnamefont
  {Montavon}}, \bibinfo {author} {\bibfnamefont {M.}~\bibnamefont {Rupp}},
  \bibinfo {author} {\bibfnamefont {V.}~\bibnamefont {Gobre}}, \bibinfo
  {author} {\bibfnamefont {A.}~\bibnamefont {Vazquez-Mayagoitia}}, \bibinfo
  {author} {\bibfnamefont {K.}~\bibnamefont {Hansen}}, \bibinfo {author}
  {\bibfnamefont {A.}~\bibnamefont {Tkatchenko}}, \bibinfo {author}
  {\bibfnamefont {K.-R.}\ \bibnamefont {Müller}}, \ and\ \bibinfo {author}
  {\bibfnamefont {O.~A.}\ \bibnamefont {von Lilienfeld}},\ }\href@noop {}
  {\bibfield  {journal} {\bibinfo  {journal} {New Journal of Physics}\ }\textbf
  {\bibinfo {volume} {15}},\ \bibinfo {pages} {095003} (\bibinfo {year}
  {2013})}\BibitemShut {NoStop}%
\bibitem [{\citenamefont {Rupp}\ \emph {et~al.}(2012)\citenamefont {Rupp},
  \citenamefont {Tkatchenko}, \citenamefont {M\"uller},\ and\ \citenamefont
  {von Lilienfeld}}]{rupp2012}%
  \BibitemOpen
  \bibfield  {author} {\bibinfo {author} {\bibfnamefont {M.}~\bibnamefont
  {Rupp}}, \bibinfo {author} {\bibfnamefont {A.}~\bibnamefont {Tkatchenko}},
  \bibinfo {author} {\bibfnamefont {K.-R.}\ \bibnamefont {M\"uller}}, \ and\
  \bibinfo {author} {\bibfnamefont {O.~A.}\ \bibnamefont {von Lilienfeld}},\
  }\href@noop {} {\bibfield  {journal} {\bibinfo  {journal} {Phys. Rev. Lett.}\
  }\textbf {\bibinfo {volume} {108}},\ \bibinfo {pages} {058301} (\bibinfo
  {year} {2012})}\BibitemShut {NoStop}%
\bibitem [{\citenamefont {Sch\"utt}\ \emph {et~al.}(2014)\citenamefont
  {Sch\"utt}, \citenamefont {Glawe}, \citenamefont {Brockherde}, \citenamefont
  {Sanna}, \citenamefont {M\"uller},\ and\ \citenamefont
  {Gross}}]{schuett2014}%
  \BibitemOpen
  \bibfield  {author} {\bibinfo {author} {\bibfnamefont {K.~T.}\ \bibnamefont
  {Sch\"utt}}, \bibinfo {author} {\bibfnamefont {H.}~\bibnamefont {Glawe}},
  \bibinfo {author} {\bibfnamefont {F.}~\bibnamefont {Brockherde}}, \bibinfo
  {author} {\bibfnamefont {A.}~\bibnamefont {Sanna}}, \bibinfo {author}
  {\bibfnamefont {K.~R.}\ \bibnamefont {M\"uller}}, \ and\ \bibinfo {author}
  {\bibfnamefont {E.~K.~U.}\ \bibnamefont {Gross}},\ }\href@noop {} {\bibfield
  {journal} {\bibinfo  {journal} {Phys. Rev. B}\ }\textbf {\bibinfo {volume}
  {89}},\ \bibinfo {pages} {205118} (\bibinfo {year} {2014})}\BibitemShut
  {NoStop}%
\bibitem [{\citenamefont {Wang}(2016)}]{wang2016}%
  \BibitemOpen
  \bibfield  {author} {\bibinfo {author} {\bibfnamefont {L.}~\bibnamefont
  {Wang}},\ }\href@noop {} {\bibfield  {journal} {\bibinfo  {journal} {Phys.
  Rev. B}\ }\textbf {\bibinfo {volume} {94}},\ \bibinfo {pages} {195105}
  (\bibinfo {year} {2016})}\BibitemShut {NoStop}%
\bibitem [{\citenamefont {{J.~Carrasquilla,
  R.~Melko}}(2017)}]{carrasquilla2017}%
  \BibitemOpen
  \bibfield  {author} {\bibinfo {author} {\bibnamefont {{J.~Carrasquilla,
  R.~Melko}}},\ }\href@noop {} {\bibfield  {journal} {\bibinfo  {journal}
  {Nature Phys}\ }\textbf {\bibinfo {volume} {13}},\ \bibinfo {pages}
  {431–434} (\bibinfo {year} {2017})}\BibitemShut {NoStop}%
\bibitem [{\citenamefont {van Nieuwenburg}\ \emph {et~al.}(2018)\citenamefont
  {van Nieuwenburg}, \citenamefont {Bairey},\ and\ \citenamefont
  {Refael}}]{vanNieuwenburg2018}%
  \BibitemOpen
  \bibfield  {author} {\bibinfo {author} {\bibfnamefont {E.}~\bibnamefont {van
  Nieuwenburg}}, \bibinfo {author} {\bibfnamefont {E.}~\bibnamefont {Bairey}},
  \ and\ \bibinfo {author} {\bibfnamefont {G.}~\bibnamefont {Refael}},\
  }\href@noop {} {\bibfield  {journal} {\bibinfo  {journal} {Phys. Rev. B}\
  }\textbf {\bibinfo {volume} {98}},\ \bibinfo {pages} {060301} (\bibinfo
  {year} {2018})}\BibitemShut {NoStop}%
\bibitem [{\citenamefont {Behler}\ and\ \citenamefont
  {Parrinello}(2007)}]{behler2007}%
  \BibitemOpen
  \bibfield  {author} {\bibinfo {author} {\bibfnamefont {J.}~\bibnamefont
  {Behler}}\ and\ \bibinfo {author} {\bibfnamefont {M.}~\bibnamefont
  {Parrinello}},\ }\href@noop {} {\bibfield  {journal} {\bibinfo  {journal}
  {Phys. Rev. Lett.}\ }\textbf {\bibinfo {volume} {98}},\ \bibinfo {pages}
  {146401} (\bibinfo {year} {2007})}\BibitemShut {NoStop}%
\bibitem [{\citenamefont {Bart\'ok}\ \emph {et~al.}(2010)\citenamefont
  {Bart\'ok}, \citenamefont {Payne}, \citenamefont {Kondor},\ and\
  \citenamefont {Cs\'anyi}}]{bartok2010}%
  \BibitemOpen
  \bibfield  {author} {\bibinfo {author} {\bibfnamefont {A.~P.}\ \bibnamefont
  {Bart\'ok}}, \bibinfo {author} {\bibfnamefont {M.~C.}\ \bibnamefont {Payne}},
  \bibinfo {author} {\bibfnamefont {R.}~\bibnamefont {Kondor}}, \ and\ \bibinfo
  {author} {\bibfnamefont {G.}~\bibnamefont {Cs\'anyi}},\ }\href@noop {}
  {\bibfield  {journal} {\bibinfo  {journal} {Phys. Rev. Lett.}\ }\textbf
  {\bibinfo {volume} {104}},\ \bibinfo {pages} {136403} (\bibinfo {year}
  {2010})}\BibitemShut {NoStop}%
\bibitem [{\citenamefont {Thompson}\ \emph {et~al.}(2015)\citenamefont
  {Thompson}, \citenamefont {Swiler}, \citenamefont {Trott}, \citenamefont
  {Foiles},\ and\ \citenamefont {Tucker}}]{thompson2015}%
  \BibitemOpen
  \bibfield  {author} {\bibinfo {author} {\bibfnamefont {A.~P.}\ \bibnamefont
  {Thompson}}, \bibinfo {author} {\bibfnamefont {L.~P.}\ \bibnamefont
  {Swiler}}, \bibinfo {author} {\bibfnamefont {C.~R.}\ \bibnamefont {Trott}},
  \bibinfo {author} {\bibfnamefont {S.~M.}\ \bibnamefont {Foiles}}, \ and\
  \bibinfo {author} {\bibfnamefont {G.~J.}\ \bibnamefont {Tucker}},\
  }\href@noop {} {\bibfield  {journal} {\bibinfo  {journal} {Journal of
  Computational Physics}\ }\textbf {\bibinfo {volume} {285}},\ \bibinfo {pages}
  {316} (\bibinfo {year} {2015})}\BibitemShut {NoStop}%
\bibitem [{\citenamefont {Shapeev}(2016)}]{shapeev2016}%
  \BibitemOpen
  \bibfield  {author} {\bibinfo {author} {\bibfnamefont {A.~V.}\ \bibnamefont
  {Shapeev}},\ }\href@noop {} {\bibfield  {journal} {\bibinfo  {journal}
  {Multiscale Modeling \& Simulation}\ }\textbf {\bibinfo {volume} {14}},\
  \bibinfo {pages} {1153} (\bibinfo {year} {2016})}\BibitemShut {NoStop}%
\bibitem [{\citenamefont {Behler}(2016)}]{behler2016}%
  \BibitemOpen
  \bibfield  {author} {\bibinfo {author} {\bibfnamefont {J.}~\bibnamefont
  {Behler}},\ }\href {\doibase 10.1063/1.4966192} {\bibfield  {journal}
  {\bibinfo  {journal} {The Journal of Chemical Physics}\ }\textbf {\bibinfo
  {volume} {145}},\ \bibinfo {pages} {170901} (\bibinfo {year} {2016})},\
  \Eprint {http://arxiv.org/abs/https://doi.org/10.1063/1.4966192}
  {https://doi.org/10.1063/1.4966192} \BibitemShut {NoStop}%
\bibitem [{\citenamefont {Hermann}\ \emph {et~al.}(2020)\citenamefont
  {Hermann}, \citenamefont {Schätzle},\ and\ \citenamefont
  {Noé}}]{hermann2020}%
  \BibitemOpen
  \bibfield  {author} {\bibinfo {author} {\bibfnamefont {J.}~\bibnamefont
  {Hermann}}, \bibinfo {author} {\bibfnamefont {Z.}~\bibnamefont {Schätzle}},
  \ and\ \bibinfo {author} {\bibfnamefont {F.}~\bibnamefont {Noé}},\
  }\href@noop {} {\bibfield  {journal} {\bibinfo  {journal} {Nature Chemistry}\
  }\textbf {\bibinfo {volume} {12}},\ \bibinfo {pages} {891} (\bibinfo {year}
  {2020})}\BibitemShut {NoStop}%
\bibitem [{\citenamefont {Manzhos}(2020)}]{manzhos2020}%
  \BibitemOpen
  \bibfield  {author} {\bibinfo {author} {\bibfnamefont {S.}~\bibnamefont
  {Manzhos}},\ }\href@noop {} {\bibfield  {journal} {\bibinfo  {journal}
  {Machine Learning: Science and Technology}\ }\textbf {\bibinfo {volume}
  {1}},\ \bibinfo {pages} {013002} (\bibinfo {year} {2020})}\BibitemShut
  {NoStop}%
\bibitem [{\citenamefont {Ghiringhelli}\ \emph {et~al.}(2015)\citenamefont
  {Ghiringhelli}, \citenamefont {Vybiral}, \citenamefont {Levchenko},
  \citenamefont {Draxl},\ and\ \citenamefont {Scheffler}}]{ghiringhelli2015}%
  \BibitemOpen
  \bibfield  {author} {\bibinfo {author} {\bibfnamefont {L.~M.}\ \bibnamefont
  {Ghiringhelli}}, \bibinfo {author} {\bibfnamefont {J.}~\bibnamefont
  {Vybiral}}, \bibinfo {author} {\bibfnamefont {S.~V.}\ \bibnamefont
  {Levchenko}}, \bibinfo {author} {\bibfnamefont {C.}~\bibnamefont {Draxl}}, \
  and\ \bibinfo {author} {\bibfnamefont {M.}~\bibnamefont {Scheffler}},\
  }\href@noop {} {\bibfield  {journal} {\bibinfo  {journal} {Phys. Rev. Lett.}\
  }\textbf {\bibinfo {volume} {114}},\ \bibinfo {pages} {105503} (\bibinfo
  {year} {2015})}\BibitemShut {NoStop}%
\bibitem [{\citenamefont {Bart\'ok}\ \emph {et~al.}(2013)\citenamefont
  {Bart\'ok}, \citenamefont {Kondor},\ and\ \citenamefont
  {Cs\'anyi}}]{bartok2013}%
  \BibitemOpen
  \bibfield  {author} {\bibinfo {author} {\bibfnamefont {A.~P.}\ \bibnamefont
  {Bart\'ok}}, \bibinfo {author} {\bibfnamefont {R.}~\bibnamefont {Kondor}}, \
  and\ \bibinfo {author} {\bibfnamefont {G.}~\bibnamefont {Cs\'anyi}},\
  }\href@noop {} {\bibfield  {journal} {\bibinfo  {journal} {Phys. Rev. B}\
  }\textbf {\bibinfo {volume} {87}},\ \bibinfo {pages} {184115} (\bibinfo
  {year} {2013})}\BibitemShut {NoStop}%
\bibitem [{\citenamefont {Rossi}\ and\ \citenamefont
  {Cumby}(2020)}]{rossi2020}%
  \BibitemOpen
  \bibfield  {author} {\bibinfo {author} {\bibfnamefont {K.}~\bibnamefont
  {Rossi}}\ and\ \bibinfo {author} {\bibfnamefont {J.}~\bibnamefont {Cumby}},\
  }\href@noop {} {\bibfield  {journal} {\bibinfo  {journal} {International
  Journal of Quantum Chemistry}\ }\textbf {\bibinfo {volume} {120}},\ \bibinfo
  {pages} {e26151} (\bibinfo {year} {2020})}\BibitemShut {NoStop}%
\bibitem [{\citenamefont {Musil}\ \emph {et~al.}(2021)\citenamefont {Musil},
  \citenamefont {Grisafi}, \citenamefont {Bartók}, \citenamefont {Ortner},
  \citenamefont {Csányi},\ and\ \citenamefont {Ceriotti}}]{musil2021}%
  \BibitemOpen
  \bibfield  {author} {\bibinfo {author} {\bibfnamefont {F.}~\bibnamefont
  {Musil}}, \bibinfo {author} {\bibfnamefont {A.}~\bibnamefont {Grisafi}},
  \bibinfo {author} {\bibfnamefont {A.~P.}\ \bibnamefont {Bartók}}, \bibinfo
  {author} {\bibfnamefont {C.}~\bibnamefont {Ortner}}, \bibinfo {author}
  {\bibfnamefont {G.}~\bibnamefont {Csányi}}, \ and\ \bibinfo {author}
  {\bibfnamefont {M.}~\bibnamefont {Ceriotti}},\ }\href@noop {} {\bibfield
  {journal} {\bibinfo  {journal} {Chemical Reviews}\ }\textbf {\bibinfo
  {volume} {121}},\ \bibinfo {pages} {9759} (\bibinfo {year}
  {2021})}\BibitemShut {NoStop}%
\bibitem [{\citenamefont {Gross}(1996)}]{gross1996}%
  \BibitemOpen
  \bibfield  {author} {\bibinfo {author} {\bibfnamefont {D.~J.}\ \bibnamefont
  {Gross}},\ }\href@noop {} {\bibfield  {journal} {\bibinfo  {journal} {Proc.
  Natl. Acad. Sci USA}\ }\textbf {\bibinfo {volume} {93}},\ \bibinfo {pages}
  {14256} (\bibinfo {year} {1996})}\BibitemShut {NoStop}%
\bibitem [{\citenamefont {Hohenberg}\ and\ \citenamefont
  {Kohn}(1964)}]{hohenberg1964}%
  \BibitemOpen
  \bibfield  {author} {\bibinfo {author} {\bibfnamefont {P.}~\bibnamefont
  {Hohenberg}}\ and\ \bibinfo {author} {\bibfnamefont {W.}~\bibnamefont
  {Kohn}},\ }\href@noop {} {\bibfield  {journal} {\bibinfo  {journal} {Phys.
  Rev.}\ }\textbf {\bibinfo {volume} {136}},\ \bibinfo {pages} {B864} (\bibinfo
  {year} {1964})}\BibitemShut {NoStop}%
\bibitem [{\citenamefont {von Lilienfeld}(2013)}]{lilienfeld2013}%
  \BibitemOpen
  \bibfield  {author} {\bibinfo {author} {\bibfnamefont {O.~A.}\ \bibnamefont
  {von Lilienfeld}},\ }\href@noop {} {\bibfield  {journal} {\bibinfo  {journal}
  {International Journal of Quantum Chemistry}\ }\textbf {\bibinfo {volume}
  {113}},\ \bibinfo {pages} {1676} (\bibinfo {year} {2013})}\BibitemShut
  {NoStop}%
\bibitem [{\citenamefont {Mills}\ \emph {et~al.}(2018)\citenamefont {Mills},
  \citenamefont {Spanner},\ and\ \citenamefont {Tamblyn}}]{mills2018}%
  \BibitemOpen
  \bibfield  {author} {\bibinfo {author} {\bibfnamefont {K.}~\bibnamefont
  {Mills}}, \bibinfo {author} {\bibfnamefont {M.}~\bibnamefont {Spanner}}, \
  and\ \bibinfo {author} {\bibfnamefont {I.}~\bibnamefont {Tamblyn}},\
  }\href@noop {} {\bibfield  {journal} {\bibinfo  {journal} {Phys. Rev. A}\
  }\textbf {\bibinfo {volume} {97}} (\bibinfo {year} {2018})}\BibitemShut
  {NoStop}%
\bibitem [{\citenamefont {Yarotsky}(2018)}]{yarotsky2018}%
  \BibitemOpen
  \bibfield  {author} {\bibinfo {author} {\bibfnamefont {D.}~\bibnamefont
  {Yarotsky}},\ }\href {http://arxiv.org/abs/1804.10306} {\bibfield  {journal}
  {\bibinfo  {journal} {CoRR}\ }\textbf {\bibinfo {volume} {abs/1804.10306}}
  (\bibinfo {year} {2018})},\ \Eprint {http://arxiv.org/abs/1804.10306}
  {arXiv:1804.10306} \BibitemShut {NoStop}%
\bibitem [{\citenamefont {Hubel}\ and\ \citenamefont
  {Wiesel}(1962)}]{hubel1962}%
  \BibitemOpen
  \bibfield  {author} {\bibinfo {author} {\bibfnamefont {D.}~\bibnamefont
  {Hubel}}\ and\ \bibinfo {author} {\bibfnamefont {T.~N.}\ \bibnamefont
  {Wiesel}},\ }\href@noop {} {\bibfield  {journal} {\bibinfo  {journal} {J
  Physiol.}\ }\textbf {\bibinfo {volume} {160}},\ \bibinfo {pages} {106}
  (\bibinfo {year} {1962})}\BibitemShut {NoStop}%
\bibitem [{\citenamefont {Fukushima}(1980)}]{fukushima1980}%
  \BibitemOpen
  \bibfield  {author} {\bibinfo {author} {\bibfnamefont {K.}~\bibnamefont
  {Fukushima}},\ }\href@noop {} {\bibfield  {journal} {\bibinfo  {journal}
  {Biological Cybernetics}\ }\textbf {\bibinfo {volume} {36}},\ \bibinfo
  {pages} {193–202} (\bibinfo {year} {1980})}\BibitemShut {NoStop}%
\bibitem [{\citenamefont {{Lecun}}\ \emph {et~al.}(1998)\citenamefont
  {{Lecun}}, \citenamefont {{Bottou}}, \citenamefont {{Bengio}},\ and\
  \citenamefont {{Haffner}}}]{lecun1998}%
  \BibitemOpen
  \bibfield  {author} {\bibinfo {author} {\bibfnamefont {Y.}~\bibnamefont
  {{Lecun}}}, \bibinfo {author} {\bibfnamefont {L.}~\bibnamefont {{Bottou}}},
  \bibinfo {author} {\bibfnamefont {Y.}~\bibnamefont {{Bengio}}}, \ and\
  \bibinfo {author} {\bibfnamefont {P.}~\bibnamefont {{Haffner}}},\ }\href
  {\doibase 10.1109/5.726791} {\bibfield  {journal} {\bibinfo  {journal}
  {Proceedings of the IEEE}\ }\textbf {\bibinfo {volume} {86}},\ \bibinfo
  {pages} {2278} (\bibinfo {year} {1998})}\BibitemShut {NoStop}%
\bibitem [{\citenamefont {Krizhevsky}\ \emph {et~al.}(2012)\citenamefont
  {Krizhevsky}, \citenamefont {Sutskever},\ and\ \citenamefont
  {Hinton}}]{krizhevsky2012}%
  \BibitemOpen
  \bibfield  {author} {\bibinfo {author} {\bibfnamefont {A.}~\bibnamefont
  {Krizhevsky}}, \bibinfo {author} {\bibfnamefont {I.}~\bibnamefont
  {Sutskever}}, \ and\ \bibinfo {author} {\bibfnamefont {G.~E.}\ \bibnamefont
  {Hinton}},\ }in\ \href
  {https://proceedings.neurips.cc/paper/2012/file/c399862d3b9d6b76c8436e924a68c45b-Paper.pdf}
  {\emph {\bibinfo {booktitle} {Advances in Neural Information Processing
  Systems}}},\ Vol.~\bibinfo {volume} {25},\ \bibinfo {editor} {edited by\
  \bibinfo {editor} {\bibfnamefont {F.}~\bibnamefont {Pereira}}, \bibinfo
  {editor} {\bibfnamefont {C.~J.~C.}\ \bibnamefont {Burges}}, \bibinfo {editor}
  {\bibfnamefont {L.}~\bibnamefont {Bottou}}, \ and\ \bibinfo {editor}
  {\bibfnamefont {K.~Q.}\ \bibnamefont {Weinberger}}}\ (\bibinfo  {publisher}
  {Curran Associates, Inc.},\ \bibinfo {year} {2012})\ pp.\ \bibinfo {pages}
  {1097--1105}\BibitemShut {NoStop}%
\bibitem [{\citenamefont {{Szegedy}}\ \emph {et~al.}(2015)\citenamefont
  {{Szegedy}}, \citenamefont {{Wei Liu}}, \citenamefont {{Yangqing Jia}},
  \citenamefont {{Sermanet}}, \citenamefont {{Reed}}, \citenamefont
  {{Anguelov}}, \citenamefont {{Erhan}}, \citenamefont {{Vanhoucke}},\ and\
  \citenamefont {{Rabinovich}}}]{szegedy2015}%
  \BibitemOpen
  \bibfield  {author} {\bibinfo {author} {\bibfnamefont {C.}~\bibnamefont
  {{Szegedy}}}, \bibinfo {author} {\bibnamefont {{Wei Liu}}}, \bibinfo {author}
  {\bibnamefont {{Yangqing Jia}}}, \bibinfo {author} {\bibfnamefont
  {P.}~\bibnamefont {{Sermanet}}}, \bibinfo {author} {\bibfnamefont
  {S.}~\bibnamefont {{Reed}}}, \bibinfo {author} {\bibfnamefont
  {D.}~\bibnamefont {{Anguelov}}}, \bibinfo {author} {\bibfnamefont
  {D.}~\bibnamefont {{Erhan}}}, \bibinfo {author} {\bibfnamefont
  {V.}~\bibnamefont {{Vanhoucke}}}, \ and\ \bibinfo {author} {\bibfnamefont
  {A.}~\bibnamefont {{Rabinovich}}},\ }in\ \href {\doibase
  10.1109/CVPR.2015.7298594} {\emph {\bibinfo {booktitle} {2015 IEEE Conference
  on Computer Vision and Pattern Recognition (CVPR)}}}\ (\bibinfo {year}
  {2015})\ pp.\ \bibinfo {pages} {1--9}\BibitemShut {NoStop}%
\bibitem [{\citenamefont {Silver}\ \emph {et~al.}(2016)\citenamefont {Silver},
  \citenamefont {Huang}, \citenamefont {Maddison}, \citenamefont {Guez},
  \citenamefont {Sifre}, \citenamefont {van~den Driessche}, \citenamefont
  {Schrittwieser}, \citenamefont {Antonoglou}, \citenamefont {Panneershelvam},
  \citenamefont {Lanctot}, \citenamefont {Dieleman}, \citenamefont {Grewe},
  \citenamefont {Nham}, \citenamefont {Kalchbrenner}, \citenamefont
  {Sutskever}, \citenamefont {Lillicrap}, \citenamefont {Leach}, \citenamefont
  {Kavukcuoglu}, \citenamefont {Graepel},\ and\ \citenamefont
  {Hassabis}}]{silver2016}%
  \BibitemOpen
  \bibfield  {author} {\bibinfo {author} {\bibfnamefont {D.}~\bibnamefont
  {Silver}}, \bibinfo {author} {\bibfnamefont {A.}~\bibnamefont {Huang}},
  \bibinfo {author} {\bibfnamefont {C.~J.}\ \bibnamefont {Maddison}}, \bibinfo
  {author} {\bibfnamefont {A.}~\bibnamefont {Guez}}, \bibinfo {author}
  {\bibfnamefont {L.}~\bibnamefont {Sifre}}, \bibinfo {author} {\bibfnamefont
  {G.}~\bibnamefont {van~den Driessche}}, \bibinfo {author} {\bibfnamefont
  {J.}~\bibnamefont {Schrittwieser}}, \bibinfo {author} {\bibfnamefont
  {I.}~\bibnamefont {Antonoglou}}, \bibinfo {author} {\bibfnamefont
  {V.}~\bibnamefont {Panneershelvam}}, \bibinfo {author} {\bibfnamefont
  {M.}~\bibnamefont {Lanctot}}, \bibinfo {author} {\bibfnamefont
  {S.}~\bibnamefont {Dieleman}}, \bibinfo {author} {\bibfnamefont
  {D.}~\bibnamefont {Grewe}}, \bibinfo {author} {\bibfnamefont
  {J.}~\bibnamefont {Nham}}, \bibinfo {author} {\bibfnamefont {N.}~\bibnamefont
  {Kalchbrenner}}, \bibinfo {author} {\bibfnamefont {I.}~\bibnamefont
  {Sutskever}}, \bibinfo {author} {\bibfnamefont {T.}~\bibnamefont
  {Lillicrap}}, \bibinfo {author} {\bibfnamefont {M.}~\bibnamefont {Leach}},
  \bibinfo {author} {\bibfnamefont {K.}~\bibnamefont {Kavukcuoglu}}, \bibinfo
  {author} {\bibfnamefont {T.}~\bibnamefont {Graepel}}, \ and\ \bibinfo
  {author} {\bibfnamefont {D.}~\bibnamefont {Hassabis}},\ }\href@noop {}
  {\bibfield  {journal} {\bibinfo  {journal} {Nature}\ }\textbf {\bibinfo
  {volume} {529}},\ \bibinfo {pages} {489} (\bibinfo {year}
  {2016})}\BibitemShut {NoStop}%
\bibitem [{\citenamefont {Mallat}(2016)}]{mallat2016}%
  \BibitemOpen
  \bibfield  {author} {\bibinfo {author} {\bibfnamefont {S.}~\bibnamefont
  {Mallat}},\ }\href@noop {} {\bibfield  {journal} {\bibinfo  {journal}
  {Philosophical Transactions of the Royal Society A: Mathematical, Physical
  and Engineering Sciences}\ }\textbf {\bibinfo {volume} {374}},\ \bibinfo
  {pages} {20150203} (\bibinfo {year} {2016})}\BibitemShut {NoStop}%
\bibitem [{\citenamefont {Lin}\ \emph {et~al.}(2017)\citenamefont {Lin},
  \citenamefont {Tegmark},\ and\ \citenamefont {Maddison}}]{lin2017}%
  \BibitemOpen
  \bibfield  {author} {\bibinfo {author} {\bibfnamefont {H.~W.}\ \bibnamefont
  {Lin}}, \bibinfo {author} {\bibfnamefont {M.}~\bibnamefont {Tegmark}}, \ and\
  \bibinfo {author} {\bibfnamefont {D.}~\bibnamefont {Maddison}, \bibfnamefont
  {Rolnick}},\ }\href@noop {} {\bibfield  {journal} {\bibinfo  {journal}
  {Journal of Statistical Physics}\ }\textbf {\bibinfo {volume} {168}},\
  \bibinfo {pages} {1223} (\bibinfo {year} {2017})}\BibitemShut {NoStop}%
\bibitem [{\citenamefont {Ashcroft}\ and\ \citenamefont
  {Mermin}(1976)}]{ashcroft1976}%
  \BibitemOpen
  \bibfield  {author} {\bibinfo {author} {\bibfnamefont {N.~W.}\ \bibnamefont
  {Ashcroft}}\ and\ \bibinfo {author} {\bibfnamefont {M.~D.}\ \bibnamefont
  {Mermin}},\ }\href@noop {} {\emph {\bibinfo {title} {{Solid State
  Physics}}}}\ (\bibinfo  {publisher} {Saunders College},\ \bibinfo {address}
  {Philadelphia},\ \bibinfo {year} {1976})\BibitemShut {NoStop}%
\bibitem [{\citenamefont {Meyer}\ and\ \citenamefont
  {Bartoli}(1981)}]{meyer1981}%
  \BibitemOpen
  \bibfield  {author} {\bibinfo {author} {\bibfnamefont {J.~R.}\ \bibnamefont
  {Meyer}}\ and\ \bibinfo {author} {\bibfnamefont {F.~J.}\ \bibnamefont
  {Bartoli}},\ }\href@noop {} {\bibfield  {journal} {\bibinfo  {journal} {Phys.
  Rev. B}\ }\textbf {\bibinfo {volume} {23}},\ \bibinfo {pages} {5413}
  (\bibinfo {year} {1981})}\BibitemShut {NoStop}%
\bibitem [{\citenamefont {Giuliani}\ and\ \citenamefont
  {Vignale}(2005)}]{giuliani2005}%
  \BibitemOpen
  \bibfield  {author} {\bibinfo {author} {\bibfnamefont {G.}~\bibnamefont
  {Giuliani}}\ and\ \bibinfo {author} {\bibfnamefont {G.}~\bibnamefont
  {Vignale}},\ }\href@noop {} {\emph {\bibinfo {title} {{Quantum Theory of
  Electron Liquid}}}}\ (\bibinfo  {publisher} {Cambridge University Press},\
  \bibinfo {address} {Cambridge, UK},\ \bibinfo {year} {2005})\BibitemShut
  {NoStop}%
\bibitem [{\citenamefont {Pohv}\ \emph {et~al.}(2002)\citenamefont {Pohv},
  \citenamefont {Rith}, \citenamefont {Scholz},\ and\ \citenamefont
  {Zetsche}}]{povh2002}%
  \BibitemOpen
  \bibfield  {author} {\bibinfo {author} {\bibfnamefont {B.}~\bibnamefont
  {Pohv}}, \bibinfo {author} {\bibfnamefont {K.}~\bibnamefont {Rith}}, \bibinfo
  {author} {\bibfnamefont {C.}~\bibnamefont {Scholz}}, \ and\ \bibinfo {author}
  {\bibfnamefont {F.}~\bibnamefont {Zetsche}},\ }\href@noop {} {\emph {\bibinfo
  {title} {{Particles and Nuclei}}}},\ \bibinfo {edition} {1st}\ ed.\ (\bibinfo
   {publisher} {Springer},\ \bibinfo {address} {Berlin},\ \bibinfo {year}
  {2002})\BibitemShut {NoStop}%
\bibitem [{\citenamefont {Bali}\ \emph {et~al.}(1967)\citenamefont {Bali},
  \citenamefont {Chu}, \citenamefont {Haymaker},\ and\ \citenamefont
  {Tan}}]{bali1967}%
  \BibitemOpen
  \bibfield  {author} {\bibinfo {author} {\bibfnamefont {N.~F.}\ \bibnamefont
  {Bali}}, \bibinfo {author} {\bibfnamefont {S.-Y.}\ \bibnamefont {Chu}},
  \bibinfo {author} {\bibfnamefont {R.~W.}\ \bibnamefont {Haymaker}}, \ and\
  \bibinfo {author} {\bibfnamefont {C.-I.}\ \bibnamefont {Tan}},\ }\href@noop
  {} {\bibfield  {journal} {\bibinfo  {journal} {Phys. Rev.}\ }\textbf
  {\bibinfo {volume} {161}},\ \bibinfo {pages} {1450} (\bibinfo {year}
  {1967})}\BibitemShut {NoStop}%
\bibitem [{\citenamefont {Nagy}\ and\ \citenamefont {Apagyi}(1998)}]{nagy1998}%
  \BibitemOpen
  \bibfield  {author} {\bibinfo {author} {\bibfnamefont {I.}~\bibnamefont
  {Nagy}}\ and\ \bibinfo {author} {\bibfnamefont {B.}~\bibnamefont {Apagyi}},\
  }\href@noop {} {\bibfield  {journal} {\bibinfo  {journal} {Phys. Rev. A}\
  }\textbf {\bibinfo {volume} {58}},\ \bibinfo {pages} {R1653} (\bibinfo {year}
  {1998})}\BibitemShut {NoStop}%
\bibitem [{\citenamefont {Lam}\ and\ \citenamefont {Varshni}(1972)}]{lam1972}%
  \BibitemOpen
  \bibfield  {author} {\bibinfo {author} {\bibfnamefont {C.~S.}\ \bibnamefont
  {Lam}}\ and\ \bibinfo {author} {\bibfnamefont {Y.~P.}\ \bibnamefont
  {Varshni}},\ }\href {\doibase 10.1103/PhysRevA.6.1391} {\bibfield  {journal}
  {\bibinfo  {journal} {Phys. Rev. A}\ }\textbf {\bibinfo {volume} {6}},\
  \bibinfo {pages} {1391} (\bibinfo {year} {1972})}\BibitemShut {NoStop}%
\bibitem [{\citenamefont {Shukla}\ and\ \citenamefont
  {Eliasson}(2008)}]{shukla2008}%
  \BibitemOpen
  \bibfield  {author} {\bibinfo {author} {\bibfnamefont {P.~K.}\ \bibnamefont
  {Shukla}}\ and\ \bibinfo {author} {\bibfnamefont {B.}~\bibnamefont
  {Eliasson}},\ }\href@noop {} {\bibfield  {journal} {\bibinfo  {journal}
  {Physics Letters A}\ }\textbf {\bibinfo {volume} {372}},\ \bibinfo {pages}
  {2897 } (\bibinfo {year} {2008})}\BibitemShut {NoStop}%
\bibitem [{\citenamefont {Shukla}\ and\ \citenamefont
  {Eliasson}(2012)}]{shukla2012}%
  \BibitemOpen
  \bibfield  {author} {\bibinfo {author} {\bibfnamefont {P.~K.}\ \bibnamefont
  {Shukla}}\ and\ \bibinfo {author} {\bibfnamefont {B.}~\bibnamefont
  {Eliasson}},\ }\href@noop {} {\bibfield  {journal} {\bibinfo  {journal}
  {Phys. Rev. Lett.}\ }\textbf {\bibinfo {volume} {108}},\ \bibinfo {pages}
  {165007} (\bibinfo {year} {2012})}\BibitemShut {NoStop}%
\bibitem [{\citenamefont {{Lin, C. Y.}}\ and\ \citenamefont {{Ho, Y.
  K.}}(2010)}]{lin2010}%
  \BibitemOpen
  \bibfield  {author} {\bibinfo {author} {\bibnamefont {{Lin, C. Y.}}}\ and\
  \bibinfo {author} {\bibnamefont {{Ho, Y. K.}}},\ }\href {\doibase
  10.1140/epjd/e2010-00009-8} {\bibfield  {journal} {\bibinfo  {journal} {Eur.
  Phys. J. D}\ }\textbf {\bibinfo {volume} {57}},\ \bibinfo {pages} {21}
  (\bibinfo {year} {2010})}\BibitemShut {NoStop}%
\bibitem [{\citenamefont {Moldabekov}\ \emph {et~al.}(2015)\citenamefont
  {Moldabekov}, \citenamefont {Schoof}, \citenamefont {Ludwig}, \citenamefont
  {Bonitz},\ and\ \citenamefont {Ramazanov}}]{moldabekov2015}%
  \BibitemOpen
  \bibfield  {author} {\bibinfo {author} {\bibfnamefont {Z.}~\bibnamefont
  {Moldabekov}}, \bibinfo {author} {\bibfnamefont {T.}~\bibnamefont {Schoof}},
  \bibinfo {author} {\bibfnamefont {P.}~\bibnamefont {Ludwig}}, \bibinfo
  {author} {\bibfnamefont {M.}~\bibnamefont {Bonitz}}, \ and\ \bibinfo {author}
  {\bibfnamefont {T.}~\bibnamefont {Ramazanov}},\ }\href@noop {} {\bibfield
  {journal} {\bibinfo  {journal} {Physics of Plasmas}\ }\textbf {\bibinfo
  {volume} {22}},\ \bibinfo {pages} {102104} (\bibinfo {year}
  {2015})}\BibitemShut {NoStop}%
\bibitem [{\citenamefont {Qi}\ \emph {et~al.}(2016)\citenamefont {Qi},
  \citenamefont {Wang},\ and\ \citenamefont {Janev}}]{qi2016}%
  \BibitemOpen
  \bibfield  {author} {\bibinfo {author} {\bibfnamefont {Y.~Y.}\ \bibnamefont
  {Qi}}, \bibinfo {author} {\bibfnamefont {J.~G.}\ \bibnamefont {Wang}}, \ and\
  \bibinfo {author} {\bibfnamefont {R.~K.}\ \bibnamefont {Janev}},\ }\href@noop
  {} {\bibfield  {journal} {\bibinfo  {journal} {Physics of Plasmas}\ }\textbf
  {\bibinfo {volume} {23}},\ \bibinfo {pages} {073302} (\bibinfo {year}
  {2016})}\BibitemShut {NoStop}%
\bibitem [{\citenamefont {Munjal}\ \emph {et~al.}(2017)\citenamefont {Munjal},
  \citenamefont {Silotia},\ and\ \citenamefont {Prasad}}]{munjal2017}%
  \BibitemOpen
  \bibfield  {author} {\bibinfo {author} {\bibfnamefont {D.}~\bibnamefont
  {Munjal}}, \bibinfo {author} {\bibfnamefont {P.}~\bibnamefont {Silotia}}, \
  and\ \bibinfo {author} {\bibfnamefont {V.}~\bibnamefont {Prasad}},\
  }\href@noop {} {\bibfield  {journal} {\bibinfo  {journal} {Physics of
  Plasmas}\ }\textbf {\bibinfo {volume} {24}},\ \bibinfo {pages} {122118}
  (\bibinfo {year} {2017})}\BibitemShut {NoStop}%
\bibitem [{\citenamefont {Krane}(1988)}]{krane1988}%
  \BibitemOpen
  \bibfield  {author} {\bibinfo {author} {\bibfnamefont {K.~S.}\ \bibnamefont
  {Krane}},\ }\href@noop {} {\emph {\bibinfo {title} {{Introductory Nuclear
  Physics}}}}\ (\bibinfo  {publisher} {John Wiley and Sons},\ \bibinfo
  {address} {U.S.A},\ \bibinfo {year} {1988})\BibitemShut {NoStop}%
\bibitem [{\citenamefont {Capri}(2002)}]{capri2002}%
  \BibitemOpen
  \bibfield  {author} {\bibinfo {author} {\bibfnamefont {A.~Z.}\ \bibnamefont
  {Capri}},\ }\href@noop {} {\emph {\bibinfo {title} {{Nonrelativistic Quantum
  Mechanics }}}},\ \bibinfo {edition} {3rd}\ ed.\ (\bibinfo  {publisher} {World
  Scientific},\ \bibinfo {address} {Singapore},\ \bibinfo {year}
  {2002})\BibitemShut {NoStop}%
\bibitem [{\citenamefont {Taylor}(1972)}]{taylor1972}%
  \BibitemOpen
  \bibfield  {author} {\bibinfo {author} {\bibfnamefont {J.~R.}\ \bibnamefont
  {Taylor}},\ }\href@noop {} {\emph {\bibinfo {title} {{Scattering Theory: The
  Quantum Theory of Nonrelativistic Collisions}}}}\ (\bibinfo  {publisher}
  {John Wiley \& Sons, Inc.},\ \bibinfo {address} {New York},\ \bibinfo {year}
  {1972})\BibitemShut {NoStop}%
\bibitem [{\citenamefont {Meijer}\ and\ \citenamefont
  {Repace}(1975)}]{meijer1975}%
  \BibitemOpen
  \bibfield  {author} {\bibinfo {author} {\bibfnamefont {P.~H.~E.}\
  \bibnamefont {Meijer}}\ and\ \bibinfo {author} {\bibfnamefont {J.~L.}\
  \bibnamefont {Repace}},\ }\href@noop {} {\bibfield  {journal} {\bibinfo
  {journal} {American Journal of Physics}\ }\textbf {\bibinfo {volume} {43}},\
  \bibinfo {pages} {428} (\bibinfo {year} {1975})}\BibitemShut {NoStop}%
\bibitem [{\citenamefont {Morse}\ and\ \citenamefont
  {Allis}(1933)}]{morse1933}%
  \BibitemOpen
  \bibfield  {author} {\bibinfo {author} {\bibfnamefont {P.~M.}\ \bibnamefont
  {Morse}}\ and\ \bibinfo {author} {\bibfnamefont {W.~P.}\ \bibnamefont
  {Allis}},\ }\href@noop {} {\bibfield  {journal} {\bibinfo  {journal} {Phys.
  Rev.}\ }\textbf {\bibinfo {volume} {44}},\ \bibinfo {pages} {269} (\bibinfo
  {year} {1933})}\BibitemShut {NoStop}%
\bibitem [{\citenamefont {Marchetti}(2019)}]{marchetti2019}%
  \BibitemOpen
  \bibfield  {author} {\bibinfo {author} {\bibfnamefont {G.}~\bibnamefont
  {Marchetti}},\ }\href@noop {} {\bibfield  {journal} {\bibinfo  {journal}
  {Journal of Applied Physics}\ }\textbf {\bibinfo {volume} {126}},\ \bibinfo
  {pages} {045713} (\bibinfo {year} {2019})}\BibitemShut {NoStop}%
\bibitem [{\citenamefont {Chadan}\ \emph {et~al.}(2001)\citenamefont {Chadan},
  \citenamefont {Kobayashi},\ and\ \citenamefont {Kobayashi}}]{chadan2001}%
  \BibitemOpen
  \bibfield  {author} {\bibinfo {author} {\bibfnamefont {K.}~\bibnamefont
  {Chadan}}, \bibinfo {author} {\bibfnamefont {R.}~\bibnamefont {Kobayashi}}, \
  and\ \bibinfo {author} {\bibfnamefont {T.}~\bibnamefont {Kobayashi}},\
  }\href@noop {} {\bibfield  {journal} {\bibinfo  {journal} {Journal of
  Mathematical Physics}\ }\textbf {\bibinfo {volume} {42}},\ \bibinfo {pages}
  {4031} (\bibinfo {year} {2001})}\BibitemShut {NoStop}%
\bibitem [{\citenamefont {Levinson}(1949)}]{levinson1949}%
  \BibitemOpen
  \bibfield  {author} {\bibinfo {author} {\bibfnamefont {N.}~\bibnamefont
  {Levinson}},\ }\href@noop {} {\bibfield  {journal} {\bibinfo  {journal} {Kgl.
  Danske Videnskab.Selskab., Mat.fys. Medd.}\ }\textbf {\bibinfo {volume} {25}}
  (\bibinfo {year} {1949})}\BibitemShut {NoStop}%
\bibitem [{\citenamefont {Portnoi}\ and\ \citenamefont
  {Galbraith}(1997)}]{portnoi1997}%
  \BibitemOpen
  \bibfield  {author} {\bibinfo {author} {\bibfnamefont {M.}~\bibnamefont
  {Portnoi}}\ and\ \bibinfo {author} {\bibfnamefont {I.}~\bibnamefont
  {Galbraith}},\ }\href@noop {} {\bibfield  {journal} {\bibinfo  {journal}
  {Solid State Communications}\ }\textbf {\bibinfo {volume} {103}},\ \bibinfo
  {pages} {325} (\bibinfo {year} {1997})}\BibitemShut {NoStop}%
\bibitem [{\citenamefont {Bethe}(1949)}]{bethe1949}%
  \BibitemOpen
  \bibfield  {author} {\bibinfo {author} {\bibfnamefont {H.~A.}\ \bibnamefont
  {Bethe}},\ }\href {\doibase 10.1103/PhysRev.76.38} {\bibfield  {journal}
  {\bibinfo  {journal} {Phys. Rev.}\ }\textbf {\bibinfo {volume} {76}},\
  \bibinfo {pages} {38} (\bibinfo {year} {1949})}\BibitemShut {NoStop}%
\bibitem [{\citenamefont {Petzold}(1983)}]{petzold1983}%
  \BibitemOpen
  \bibfield  {author} {\bibinfo {author} {\bibfnamefont {L.}~\bibnamefont
  {Petzold}},\ }\href@noop {} {\bibfield  {journal} {\bibinfo  {journal} {SIAM
  Journal on Scientific and Statistical Computing}\ }\textbf {\bibinfo {volume}
  {4}},\ \bibinfo {pages} {136} (\bibinfo {year} {1983})}\BibitemShut {NoStop}%
\bibitem [{\citenamefont {Abadi}\ \emph {et~al.}(2016)\citenamefont {Abadi},
  \citenamefont {Barham}, \citenamefont {Chen}, \citenamefont {Chen},
  \citenamefont {Davis}, \citenamefont {Dean}, \citenamefont {Devin},
  \citenamefont {Ghemawat}, \citenamefont {Irving}, \citenamefont {Isard},
  \citenamefont {Kudlur}, \citenamefont {Levenberg}, \citenamefont {Monga},
  \citenamefont {Moore}, \citenamefont {Murray}, \citenamefont {Steiner},
  \citenamefont {Tucker}, \citenamefont {Vasudevan}, \citenamefont {Warden},
  \citenamefont {Wicke}, \citenamefont {Yu},\ and\ \citenamefont
  {Zheng}}]{tensorflow}%
  \BibitemOpen
  \bibfield  {author} {\bibinfo {author} {\bibfnamefont {M.}~\bibnamefont
  {Abadi}}, \bibinfo {author} {\bibfnamefont {P.}~\bibnamefont {Barham}},
  \bibinfo {author} {\bibfnamefont {J.}~\bibnamefont {Chen}}, \bibinfo {author}
  {\bibfnamefont {Z.}~\bibnamefont {Chen}}, \bibinfo {author} {\bibfnamefont
  {A.}~\bibnamefont {Davis}}, \bibinfo {author} {\bibfnamefont
  {J.}~\bibnamefont {Dean}}, \bibinfo {author} {\bibfnamefont {M.}~\bibnamefont
  {Devin}}, \bibinfo {author} {\bibfnamefont {S.}~\bibnamefont {Ghemawat}},
  \bibinfo {author} {\bibfnamefont {G.}~\bibnamefont {Irving}}, \bibinfo
  {author} {\bibfnamefont {M.}~\bibnamefont {Isard}}, \bibinfo {author}
  {\bibfnamefont {M.}~\bibnamefont {Kudlur}}, \bibinfo {author} {\bibfnamefont
  {J.}~\bibnamefont {Levenberg}}, \bibinfo {author} {\bibfnamefont
  {R.}~\bibnamefont {Monga}}, \bibinfo {author} {\bibfnamefont
  {S.}~\bibnamefont {Moore}}, \bibinfo {author} {\bibfnamefont {D.~G.}\
  \bibnamefont {Murray}}, \bibinfo {author} {\bibfnamefont {B.}~\bibnamefont
  {Steiner}}, \bibinfo {author} {\bibfnamefont {P.}~\bibnamefont {Tucker}},
  \bibinfo {author} {\bibfnamefont {V.}~\bibnamefont {Vasudevan}}, \bibinfo
  {author} {\bibfnamefont {P.}~\bibnamefont {Warden}}, \bibinfo {author}
  {\bibfnamefont {M.}~\bibnamefont {Wicke}}, \bibinfo {author} {\bibfnamefont
  {Y.}~\bibnamefont {Yu}}, \ and\ \bibinfo {author} {\bibfnamefont
  {X.}~\bibnamefont {Zheng}},\ }in\ \href@noop {} {\emph {\bibinfo {booktitle}
  {Proceedings of the 12th USENIX Conference on Operating Systems Design and
  Implementation}}},\ \bibinfo {series and number} {OSDI'16}\ (\bibinfo
  {publisher} {USENIX Association},\ \bibinfo {address} {USA},\ \bibinfo {year}
  {2016})\ p.\ \bibinfo {pages} {265–283}\BibitemShut {NoStop}%
\bibitem [{\citenamefont {Kingma}\ and\ \citenamefont {Ba}(2014)}]{adam_opt}%
  \BibitemOpen
  \bibfield  {author} {\bibinfo {author} {\bibfnamefont {D.~P.}\ \bibnamefont
  {Kingma}}\ and\ \bibinfo {author} {\bibfnamefont {J.}~\bibnamefont {Ba}},\
  }\href {http://arxiv.org/abs/1412.6980} {\enquote {\bibinfo {title} {Adam: A
  method for stochastic optimization},}\ } (\bibinfo {year} {2014})\BibitemShut
  {NoStop}%
\bibitem [{Note1()}]{Note1}%
  \BibitemOpen
  \bibinfo {note} {Note that in this case, using the same model hyperparameters
  as in the previous cases, resulted in higher variability in validation loss
  during training. We regularized the model by reducing the learning rate of
  the Adam optimizer to $10^{-6}$ and increasing the training epochs increased
  to 5000.}\BibitemShut {Stop}%
\end{thebibliography}%

\bibliographystyle{apsrev4-1}

\clearpage

\end{document}